\documentclass[prd,aps,floatfix,nofootinbib, 10 pt]{revtex4}

\newcommand{\bea}{\begin{eqnarray}}
\newcommand{\eea}{\end{eqnarray}}
\newcommand{\beq}{\begin{eqnarray}}
\newcommand{\eeq}{\end{eqnarray}}
\newcommand{\nn}{\nonumber}

\usepackage{amsmath,graphicx,color,epsfig}

\begin{document}
\title{Asymmetries in $B\to K^{\ast}\ell^{+}\ell^{-}$ Decays and Two Higgs Doublet Model}
\author{Ishtiaq Ahmed$^{1}$}
\author{M. Jamil Aslam$^{2}$}
\author{M. Ali Paracha$^{3}$}
\affiliation{$^{1}$National Centre for Physics,\\
Quaid-i-Azam University Campus, Islamabad 45320, Pakistan }
\affiliation{$^{2}$Department of Physics,\\
Quaid-i-Azam University, Islamabad 45320, Pakistan}
\affiliation{$^{3}$Department of Physics,\\
School of Natural Sciences, NUST, Islamabad, Pakistan}
\date{\today}

\begin{abstract}
The study of polarized branching ratio and different lepton polarization asymmetries in the exclusive rare decays
$B\rightarrow K^{\ast}\ell^{+}\ell^{-}$ ($\ell=\mu , \tau$) have been made in the standard model (SM) and in type III of the THDM. It has
been found that the effects arise from the THDM are quite promising in the longitudinal and transverse polarized branching ratios $(\mathcal{B}_{L, T})$. Likewise, in case of $\mu$'s as 
final state leptons, the polarized branching ratios have an order of magnitude difference from their SM results. Similar trends have also been observed both in the longitudinal and
normal lepton polarization asymmetries $P_{L, N}$ as well as their average values. We hope that the possible signatures of these observables in $B\rightarrow K^{\ast}\ell^{+}\ell^{-}$ decays
at different on going and future experiments will serve as a good tool in the indirect searches of an extra Higgs boson doublet.
\end{abstract}
\maketitle



\section{Introduction}\label{intro}

Rare $B$ meson decays based on the flavour-changing-neutral-current (FCNC) transitions, $b \to s$, are of special interest because of their occurrence at loop level through top quark in the Standard Model (SM). Therefore, these decays may be used to determine the Cabibbo-Kobayshi-Masakawa (CKM) matrix elements $V_{tq}$ $(q=u, d, s)$. In addition, the possibility to constraint potential New Physics (NP) contributions at low energies has made $b \to s$ transitions the focus of many theoretical and experimental studies. During the first run of the LHC, we are not been able to observe the direct search of the NP. However, the run2 of the LHC might give us an opportunity to establish the nature of physics beyond the SM, i.e., the NP. At low energy frontiers the different $B$ factories and the LHCb experiment provided us with a lot of rather precise results on $b \to s$ transitions. Recently, the LHCb angular analysis of exclusive  $B \to K^{\ast} \mu^{+}\mu^{-}$  decay suggested significant deviation from the SM predictions \cite{LHCb} which are noticeable in the observables $\mathcal{P}_{5}^{\prime}$ and $\mathcal{P}_2$. In order to understand these anomalies, this exclusive decay mode has been extensively studied in literature  \cite{Matias1, Hofer, Matias2, Matias3, Matias4, Matias5, Matias6, Matias7, Matias8}.

At the low energies, the decays like $B \to K^{\ast} \gamma$ and $B \to K^{\ast} \ell^{+} \ell^{-}$ are analyzed through the effective Hamiltonian approach where the physics at short-distances is separated from the one at the long-distances. One of the most promising features of $B \to K^{\ast} \ell^{+} \ell^{-}$ decays is that they can be used to identify NP inside the short-distance Wilson coefficients, i.e., $\mathcal{C}_i = \mathcal{C}_{i}^{\text{SM}} + \mathcal{C}_{i}^{\text{NP}}$. This way of identifying the NP through Wilson coefficients not only for the SM electromagnetic and dileptonic operators
\begin{eqnarray}
O_{7} &=&\frac{e^{2}}{16\pi ^{2}}m_{b}(\bar{s}\sigma _{\mu \nu }Rb)F^{\mu
\nu },  \label{2} \\
O_{9} &=&\frac{e^{2}}{16\pi ^{2}}\left( \bar{s}\gamma _{\mu }Lb\right) \bar{\ell
}\gamma ^{\mu }\ell,  \label{3} \\
O_{10} &=&\frac{e^{2}}{16\pi ^{2}}\left( \bar{s}\gamma _{\mu }Lb\right) \bar{\ell}\gamma ^{\mu }\gamma ^{5}\ell,  \label{4}
\end{eqnarray}
with $L,R=\frac{1}{2}\left( 1\mp \gamma ^{5}\right)$ are the chirality projection operators, Additionally, there are the chirality flipped operators, i.e., 
\begin{eqnarray}
O^{\prime}_{7} &=&\frac{e^{2}}{16\pi ^{2}}m_{b}(\bar{s}\sigma _{\mu \nu }Lb)F^{\mu
\nu }, \label{2p}\\
O^{\prime}_{9} &=&\frac{e^{2}}{16\pi ^{2}}\left( \bar{s}\gamma _{\mu }Rb\right) \bar{\ell
}\gamma ^{\mu }\ell, \label{3p}\\
O^{\prime}_{10} &=&\frac{e^{2}}{16\pi ^{2}}\left( \bar{s}\gamma _{\mu }Rb\right) \bar{\ell}\gamma ^{\mu }\gamma ^{5}\ell,  \label{4p}
\end{eqnarray}
as well as for the scalar and pseudo-scalar operators: 
\begin{eqnarray}
Q_{1} &=&\frac{e^{2}}{16\pi ^{2}}\left( \bar{s}Rb\right) \bar{\ell}\ell,  \label{q1}\\
Q^{\prime}_{1} &=&\frac{e^{2}}{16\pi ^{2}}\left( \bar{s}Lb\right) \bar{\ell}\ell , \label{q1p}\\
Q_{2} &=&\frac{e^{2}}{16\pi ^{2}}\left( \bar{s}Rb\right) \bar{\ell}\gamma ^{5}\ell , \label{q2}\\
Q^{\prime}_{2} &=&\frac{e^{2}}{16\pi ^{2}}\left( \bar{s}Lb\right) \bar{\ell}\gamma ^{5}\ell. \label{q2p}
\end{eqnarray}

The focus of the present study is to  investigate the decay processes $B\rightarrow K^{\ast }\ell^{+}\ell^{-}$ in the two-Higgs-doublet-model (THDM) which is  among the natural and the most popular extensions of the SM. In SM the generation of mass is through the one Higgs doublet, whereas, in the THDM we consider two complex Higgs doublets. In order to avoid the tree level FCNC transitions in the THDM an ad-hoc discrete symmetry \cite{adhoc} is imposed which leads to two different possibilities:

\begin{itemize}
\item In order to keep the flavor conservation at the tree level if one couples all the fermions to only one of the Higgs doublet then it is called to be the model I.
\item In second version the Higgs sector of the THDM coincides with that of the supersymmetric model, i.e., when the up-type quarks are coupled to the one Higgs doublet and the down-type to the second one. This is named as model II in the literature.
\end{itemize}
The physical contents of the Higgs sector contain a pair of charged Higgs bosons $H^{\pm}$,  two neutral scalar Higgs bosons $H^{0}$, $h^{0}$ and a psudo-scalar Higgs $A^{0}$. The vacuum expectation values of the two Higgs doublets are denoted by $v_{1}$ and $v_{2}$, and the interactions of the fermions to the Higgs fields depend on the $tan\beta = v_{2}/v_{1}$ which is  a free parameter of the model.

There is another possibility where the discrete symmetry is not imposed  which in turn leads to the most general form of the THDM, i.e., to say model III. The FCNC transitions are allowed at the tree level in this version. The indirect constraints on the masses of charged Higgs bosons $m_{H^{\pm}}$, the neutral scalars $m_{H^0}, m_{h^0}$ and pseudo-scalar  $m_{A^0}$ along with the fermion Higgs interaction vertex, $tan{\beta}$ are obtained from the experimental observation of branching ratios of $b \to s \gamma$, $B \to D \tau \nu_{\tau}$ decays  and $K - \bar{K}$ and $B - \bar{B}$ mixing in the literature \cite{constraints}.  Consistent with the low energy constraints, the FCNCs involving the third generation are not as severely suppressed as the one involving the first two generations. Contrary to the SM and the first two versions of the THDMs mentioned above, here exist a single $CP$ phase of vacuum which leads to a rich source of the phenomenological studies of $CP$ violating observables \cite{constraints, constraint1, constraint2, constraint3, constraint4, Falahati}. 

Here the focus of discussion are the polarized branching ratio ${\cal BR}$ and the different lepton polarization asymmetries for $B \to K^* \ell^+\ell^-$ decay in type III of the THDM and compare them with their SM values.  In regard to the FCNC transitions, the remarkable feature of the THDM is that in this class of models the manifestation of the NP is two fold, i.e., through the modification of the Wilson coefficients as well as through the new operators in the effective Hamiltonian. Therefore, the measurement of above mentioned observables in $B \to K^* \ell^+\ell^-$ decays might help us to get hints of the profile of different parameters of the THDM in these decays.

The paper is structured as follows. In Sec. \ref{tf} we present the theoretical framework for the decay $B\rightarrow K^{\ast }\ell^{+}\ell^{-}$ necessary for the study of the THDM. In Sec. \ref{po} we present the basic formulas for physical observables such as decay rate, forward-backward asymmetries ${\cal A}_{FB}$ and the lepton polarization asymmetries. Whereas the numerical analysis and discussion on these observables is given in Sec. \ref{num}. Section \ref{con} gives the summary of the results of our study.

\section{Effective Hamiltonian}\label{tf}
In this section we give the effective Hamiltonian for $B\rightarrow K^{\ast } \ell^{+}
\ell^{-}$ decays that at quark level are governed by the transition $b\rightarrow s\ell^{+}\ell^{-}$. After integrating out the heavy degrees of freedom from the full theory, the general form of the affective Hamiltonian for the SM and the THDM can be written as \cite{Huang}:
\begin{align}
H_{eff}=-\frac{4G_{F}}{\sqrt{2}}V_{tb}V_{ts}^{\ast }&\bigg[\sum\limits_{i=1}^{10}C_{i}(\mu )O_{i}(\mu)+\sum\limits_{i=1}^{10}C_{Qi}(\mu )Q_{i}(\mu) \bigg],  \label{1}
\end{align}%
where $O_{i}(\mu )$ $(i=1,2, \cdots,10)$ are the four quark operators and $%
C_{i}(\mu )$ are the corresponding Wilson coefficients at the energy scale $%
\mu $ which is usually taken to be the $b$-quark mass $\left(
m_{b}\right) $. The theoretical uncertainties related to the renormalization
scale can be reduced when the next to leading logarithm corrections are
included. Also the contribution from the charged Higgs boson in case of the THDM is absorbed in these Wilson coefficients. The new operators $Q_{i}(i=1, 2, \cdots, 10)$ come from the NHBs exchange diagrams,
whose manifest forms and corresponding Wilson coefficients are summarized in the appendix. 
The explicit forms of the operators responsible for the decay $%
B\rightarrow K^{\ast}\ell ^{+}\ell ^{-}$, in the SM and the THDM are:
\begin{subequations}
\begin{eqnarray}
O_{7} &=&\frac{e^{2}}{16\pi ^{2}}(m_{b}-m_s)(\bar{s}\sigma _{\mu \nu }Rb)F^{\mu
\nu },  \label{2} \\
O_{9} &=&\frac{e^{2}}{16\pi ^{2}}\left( \bar{s}\gamma _{\mu }Lb\right) \bar{\ell
}\gamma ^{\mu }\ell,  \label{3} \\
O_{10} &=&\frac{e^{2}}{16\pi ^{2}}\left( \bar{s}\gamma _{\mu }Lb\right) \bar{\ell}\gamma ^{\mu }\gamma ^{5}\ell,  \label{4}\\
Q_{1} &=&\frac{e^{2}}{16\pi ^{2}}\left( \bar{s}Rb\right) \bar{\ell}\ell,  \label{q1}\\
Q_{2} &=&\frac{e^{2}}{16\pi ^{2}}\left( \bar{s}Rb\right) \bar{\ell}\gamma ^{5}\ell,  \label{q2}\\
\end{eqnarray}%
\end{subequations}
with $L,R=\frac{1}{2}\left( 1\mp \gamma ^{5}\right)$.

Using the effective Hamiltonian given in Eq. (\ref{1}) the free quark
amplitude for $b\rightarrow s\ell ^{+}\ell ^{-}$ can be written as%
\begin{eqnarray}
\mathcal{M}(b\rightarrow s\ell^{+}\ell^{-}) &=&-\frac{G_{F}\alpha }{\sqrt{2}\pi }V_{tb}V_{ts}^{\ast }\bigg[\widetilde{C}_{9}^{eff}\left( \mu
\right) (\bar{s}\gamma _{\mu }Lb)(\bar{\ell}\gamma ^{\mu }\ell)+\widetilde{C}_{10}(\bar{s}%
\gamma _{\mu }Lb)(\bar{\ell}\gamma ^{\mu }\gamma ^{5}\ell)  \notag \\
&&-2\widetilde{C}_{7}^{eff}\left( \mu \right) \frac{m_{b}}{s}(\bar{s}i\sigma _{\mu
\nu }q^{\nu }Rb)\bar{\ell}\gamma ^{\mu }\ell+C_{Q_1}\left( \bar{s}Rb\right) \left(\bar{\ell}\ell \right) \notag\\
&&+C_{Q _2}\left( \bar{s}Rb\right) \left(\bar{\ell}\gamma^{5}\ell \right)\bigg], \label{5a}
\end{eqnarray}
where $q$ is the momentum transfer. Because of the absence of  $Z$-boson in the effective theory the operator $%
O_{10}$ given in Eq.(\ref{4}) can not be induced by the insertion of four
quark operators.
Therefore, the corresponding Wilson coefficient $C_{10}$ does not renormalize under QCD
corrections and is independent of the energy scale $\mu .$ Additionally the
above quark level decay amplitude can get contributions from the matrix
element of four quark operators, $\sum_{i=1}^{6}\left\langle
\ell^{+}\ell^{-}s\left\vert O_{i}\right\vert b\right\rangle ,$ which are usually
absorbed into the effective Wilson coefficient $C_{9}^{eff}(\mu )$ and can
be written as \cite{25, 26, 27, 28, 29, 30, 31}
\begin{equation*}
C_{9}^{eff}(\mu )=C_{9}(\mu )+Y_{SD}(z,s^{\prime })+Y_{LD}(z,s^{\prime }),
\end{equation*}%
where $z=m_{c}/m_{b}$ and $s^{\prime }=q^{2}/m_{b}^{2}$. $Y_{SD}(z,s^{\prime
})$ describes the short distance contributions from four-quark operators far
away from the $c\bar{c}$ resonance regions, and this can be calculated
reliably in the perturbative theory. However the long distance contribution $%
Y_{LD}(z,s^{\prime })$ cannot be calculated by using the first principles of
QCD, so they are usually parametrized in the form of a phenomenological
Breit-Wigner formula making use of the vacuum saturation approximation and the
quark hadron duality. The expressions for the short-distance and the long-distance contributions $Y_{SD}(z,s^{\prime })$ is given as
\begin{eqnarray}
Y_{SD}(z,s^{\prime }) &=&h(z,s^{\prime })\left[3C_{1}(\mu )+C_{2}(\mu
)+3C_{3}(\mu )+C_{4}(\mu )+3C_{5}(\mu )+C_{6}(\mu )\right]  \notag \\
&&-\frac{1}{2}h(1,s^{\prime })\left[4C_{3}(\mu )+4C_{4}(\mu )+3C_{5}(\mu
)+C_{6}(\mu )\right]  \notag \\
&&-\frac{1}{2}h(0,s^{\prime })\left[C_{3}(\mu )+3C_{4}(\mu )\right]+{\frac{2}{9}}\left[3C_{3}(\mu )+C_{4}(\mu )+3C_{5}(\mu )+C_{6}(\mu )\right],
\end{eqnarray}
\begin{eqnarray}
Y_{LD}(z,s^{\prime }) &=&\frac{3}{\alpha _{em}^{2}}(3C_{1}(\mu )+C_{2}(\mu
)+3C_{3}(\mu )+C_{4}(\mu )+3C_{5}(\mu )+C_{6}(\mu ))  \notag \\
&&\times\sum_{j=\psi ,\psi ^{\prime }}\omega _{j}(q^{2})k_{j}\frac{\pi \Gamma
(j\rightarrow l^{+}l^{-})M_{j}}{q^{2}-M_{j}^{2}+iM_{j}\Gamma _{j}^{tot}},
\label{LD}
\end{eqnarray}%
with
\begin{eqnarray}
h(z,s^{\prime }) &=&-{\frac{8}{9}}\mathrm{ln}z+{\frac{8}{27}}+{\frac{4}{9}}x-%
{\frac{2}{9}}(2+x)|1-x|^{1/2}\left\{
\begin{array}{l}
\ln \left| \frac{\sqrt{1-x}+1}{\sqrt{1-x}-1}\right| -i\pi \quad \mathrm{for}{%
{\ }x\equiv 4z^{2}/s^{\prime }<1} \\
2\arctan \frac{1}{\sqrt{x-1}}\qquad \mathrm{for}{{\ }x\equiv
4z^{2}/s^{\prime }>1}%
\end{array}%
\right. ,  \notag \\
h(0,s^{\prime }) &=&{\frac{8}{27}}-{\frac{8}{9}}\mathrm{ln}{\frac{m_{b}}{\mu
}}-{\frac{4}{9}}\mathrm{ln}s^{\prime }+{\frac{4}{9}}i\pi \,\,.
\end{eqnarray}%
Here $M_{j}(\Gamma _{j}^{tot})$ are the masses (widths) of the intermediate
resonant states and $\Gamma (j\rightarrow l^{+}l^{-})$ denote the partial
decay width for the transition of vector charmonium state to massless lepton
pair, which can be expressed in terms of the decay constant of charmonium
through the relation \cite{32}
\begin{equation*}
\Gamma (j\rightarrow \ell^{+}\ell^{-})=\pi \alpha _{em}^{2}{\frac{16}{27}}{\frac{%
f_{j}^{2}}{M_{j}}}.
\end{equation*}%
The phenomenological parameter $k_{j}$ in Eq.(\ref{LD}) is to account for
inadequacies of the factorization approximation, and it can be determined
from
\begin{equation*}
{\cal BR}(B\rightarrow K^{\ast} J/\psi \rightarrow  K^{\ast}
\ell^{+}\ell^{-})={\cal BR}(B\rightarrow K^{\ast} J/\psi )\cdot {\cal BR}(J/\psi\rightarrow \ell^{+}\ell^{-}).
\end{equation*}%
The function $\omega _{j}(q^{2})$ introduced in Eq.(\ref{LD}) is to
compensate the naive treatment of long distance contributions due to the
charm quark loop in the spirit of quark-hadron duality, which can
overestimate the genuine effect of the charm quark at small $q^{2}$
remarkably \footnote{%
For a more detailed discussion on long-distance and short-distance
contributions from the charm loop, one can refer to references \cite{aali, 32, b to s 2, b to s 3,charm loop 1, charm loop 2,charm
loop 3}.}. The quantity $\omega _{j}(q^{2})$ can be normalized to $\omega
_{j}(M_{\psi _{j}}^{2})=1$, but its exact form is unknown at present. Since
the dominant contribution of the resonances is in the vicinity of the
intermediate $\psi _{i}$ masses, we will simply use $\omega _{j}(q^{2})=1$
in our numerical calculations.

It has already been pointed out that in $B \to K^* \ell^+ \ell^-$ the charm-loop pollution significantly modify the results of various asymmetries in different bins of the square of momentum transfere $(s)$. The perturbative charm-loop contribution is usually absorbed into the definition of $C_9^{eff}$ \cite{Beneke}. The long-distance contribution is difficult to estimate,
and to incorporate them a universal correction to $C_9$ arising from the long-distance charm-loop contribution, that we parametrize as \cite{Mannel, Virto}: 
\begin{equation}
\delta C^{c \bar{c},LD}_{9} =\delta_i \frac{a+bs(c-s)}{s(c-s)}, \label{cloop}
\end{equation}
with $a \in [2,7]$ GeV$^4$, $b \in [0.1,0.2]$ and $c \in [9.2, 9.5]$ GeV$^2$, where as the range of the parameter $\delta_i$ is $[-1,1]$. 

Moreover, the non factorizable effects from the charm quark loop brings further
corrections to the radiative transition $b\rightarrow s\gamma ,$ and these
can be absorbed into the effective Wilson coefficients $C_{7}^{eff}$ which
then takes the form \cite{Mannel, 32, b to s 2, b to s 3,charm loop 1, charm loop 2,charm
loop 3}
\begin{equation*}
C_{7}^{eff}(\mu )=C_{7}(\mu )+C_{b\rightarrow s\gamma }(\mu ),
\end{equation*}%
with
\begin{eqnarray}
C_{b\rightarrow s\gamma }(\mu ) &=&i\alpha _{s}\left[ \frac{2}{9}\eta
^{14/23}(G_{1}(x_{t})-0.1687)-0.03C_{2}(\mu )\right],  \label{8} \\
G_{1}(x_{t}) &=&\frac{x_{t}\left( x_{t}^{2}-5x_{t}-2\right) }{8\left(
x_{t}-1\right) ^{3}}+\frac{3x_{t}^{2}\ln ^{2}x_{t}}{4\left( x_{t}-1\right)
^{4}},  \label{9}
\end{eqnarray}%
where $\eta =\alpha _{s}(m_{W})/\alpha _{s}(\mu ),$ \ $%
x_{t}=m_{t}^{2}/m_{W}^{2}$ and $C_{b\rightarrow s\gamma }$ is the absorptive
part for the $b\rightarrow sc\bar{c}\rightarrow s\gamma $ re-scattering.

\subsection{Parameterizations of the Matrix Elements and Form Factors}\label{ff}

The exclusive $B\rightarrow K^{\ast }\ell^{+}\ell^{-}$ decay involves the
hadronic matrix elements which can be obtained by sandwiching the quark
level operators given in Eq. (\ref{5a}) between initial state $%
B$ meson and final state $K^*$ meson. These can be
parametrized in terms of form factors which are the scalar functions of the
square of the four momentum transfer($q^{2}=(p-k)^{2}).$ The non vanishing
matrix elements for the process $B\rightarrow K^{\ast }$ can be
parametrized in terms of the seven form factors as follows%
\begin{eqnarray}
\left\langle K^{\ast }(k,\varepsilon )\left\vert \bar{s}\gamma _{\mu
}b\right\vert B(p)\right\rangle &=&\frac{2A_{V}(q^{2})}{%
M_{B}+M_{K^{\ast}}}\epsilon _{\mu \nu \alpha \beta }\varepsilon
^{\ast \nu }p^{\alpha }k^{\beta },  \label{10} \\
\left\langle K^{\ast }(k,\varepsilon )\left\vert \bar{s}\gamma _{\mu
}\gamma _{5}b\right\vert B(p)\right\rangle &=&i\left(
M_{B}+M_{K^{\ast }}\right) \varepsilon ^{\ast \mu }A_{1}(q^{2})-i\frac{A_{2}\left( q^{2}\right) }{M_{B}+M_{K^{\ast }}}%
(\varepsilon ^{\ast }\cdot p)\left( p+k\right) ^{\mu }\notag\\&& -2i\frac{A_3\left( q^{2}\right)-A_0 \left( q^{2}\right)}{q^2}%
(\varepsilon ^{\ast }\cdot q)q^{\mu }  M_{K^{\ast }} , \label{11}\\
\end{eqnarray}%
where $p$ is the momentum of the $B$, $\varepsilon $ and $k$ are the
polarization vector and momentum of the final state $K^{\ast }$ vector meson. The form factor ${A}_{3}(q^{2})$ can be parametrized as
\begin{equation}
{A}_{3}(q^{2})=\frac{M_{B}+M_{K^{\ast }}}{2M_{K^{\ast }}}%
A_{1}(q^{2})-\frac{M_{B}-M_{K^{\ast }}}{2M_{K^{\ast }}}
A_{2}(q^{2}). \label{12a}
\end{equation}%

In addition to the above form factors there are some penguin form factors,
which we can write as
\begin{eqnarray}
\left\langle K^{\ast }(k,\varepsilon )\left\vert \bar{s}\sigma _{\mu \nu
}q^{\nu }b\right\vert B(p)\right\rangle &=&2iT_{1}(q^{2})\epsilon _{\mu
\nu \alpha \beta }\varepsilon ^{\ast \nu }p^{\alpha }k^{\beta },  \label{13a}
\\
\left\langle K^{\ast }(k,\varepsilon )\left\vert \bar{s}\sigma _{\mu \nu
}q^{\nu }\gamma ^{5}b\right\vert B(p)\right\rangle &=&\left[ \left(
M_{B}^{2}-M_{K^{\ast }}^{2}\right) \varepsilon _{\mu }^{\ast
}-(\varepsilon ^{\ast }\cdot p)(p+k)_{\mu }\right] T_{2}(q^{2})  \notag \\
&&  \label{13b} \\
&&+(\varepsilon ^{\ast }\cdot p)\left[ q_{\mu }-\frac{q^{2}}{%
M_{B}^{2}-M_{K^{\ast }}^{2}}(p+k)_{\mu }\right] T_{3}(q^{2}).  \notag
\end{eqnarray}%
The form factors $A_{V}\left( q^{2}\right) ,~A_{1}\left( q^{2}\right) ,$ $%
~A_{2}\left( q^{2}\right) ,~A_{3}\left( q^{2}\right) ,~T_{1}\left(
q^{2}\right) ,~T_{2}\left( q^{2}\right) ,~T_{3}\left( q^{2}\right) $ are the
non-perturbative quantities and to calculate them one has to rely on some
non-perturbative approaches. In our numerical analysis we use the form
factors calculated by using Light Cone Sum Rules (LCSR) \cite{aali}. The dependence of these
form factors on square of the momentum transfer $(q^{2})$ can be written as%
\begin{equation}
F\left( q^{2}\right) =F\left( 0\right)\exp{[a\frac{q^2}{M_B^2}+b\frac{q^2}{M_B^2}]}.  \label{ff-param}
\end{equation}%
where the values of the parameters $F\left( 0\right) $, $a$ and $b$ is given in Table \ref{formfactor}.

\begin{table}[tbh]
\centering
\begin{tabular}{cccc}
\hline\hline
 $F(q^{2})$ & $\hspace{2cm}F(0)$ & $\hspace{2cm}a$ & $\hspace{%
2cm}b$ \\ \hline
 $A_{V}\left( q^{2}\right) $ & $\hspace{2cm}0.457^{+0.091}_{-0.058}$ & $%
\hspace{2cm}1.482$ & $\hspace{2cm}1.015$ \\ \hline
 $A_{1}(q^{2})$ & $\hspace{2cm}0.337^{+0.048}_{-0.043}$ & $\hspace{2cm}%
0.602$ & $\hspace{2cm}0.258$ \\ \hline
 $A_{2}(q^{2})$ & $\hspace{2cm}0.282^{+0.038}_{-0.036}$ & $\hspace{2cm}%
1.172$ & $\hspace{2cm}0.567$ \\ \hline
 $A_{0}(q^{2})$ & $\hspace{2cm}0.471^{+0.227}_{-0.059}$ & $\hspace{2cm}%
1.505$ & $\hspace{2cm}0.710$ \\ \hline
 $T_{1}(q^{2})$ & $\hspace{2cm}0.379^{+0.058}_{-0.045}$ & $\hspace{2cm}%
1.519$ & $\hspace{2cm}1.030$ \\ \hline
 $T_{2}(q^{2})$ & $\hspace{2cm}0.379^{+0.058}_{-0.045}$ & $\hspace{2cm}%
0.517$ & $\hspace{2cm}0.426$ \\ \hline
 $T_{3}(q^{2})$ & $\hspace{2cm}0.260^{+0.035}_{-0.026}$ & $\hspace{2cm}%
1.129$ & $\hspace{2cm}1.128$ \\ \hline\hline
\end{tabular}
\caption{$B\rightarrow K^{\ast }$ form factors corresponding to penguin
contributions in the LCSR.
$F(0)$ denotes the value of form factors at $q^{2}=0$ while $a$ and $b$
are the parameters in the parameterizations shown in Eq. (\ref{ff-param}) \cite{aali}.}
\label{formfactor}
\end{table}

From Eq. (\ref{5a}) it is straightforward to write the penguin
amplitude
\begin{equation}
\mathcal{M}=-\frac{G_{F}\alpha }{2\sqrt{2}\pi }%
V_{tb}V_{ts}^{\ast }\left[ {\cal T}_{\mu }^{1}(\bar{\ell}\gamma ^{\mu }\ell)+{\cal T}_{\mu
}^{2}\left( \bar{\ell}\gamma ^{\mu }\gamma ^{5}\ell\right) +{\cal T}\left( \bar{\ell}\ell\right)\right], \label{59}
\end{equation}%
where%
\begin{eqnarray}
{\cal T}_{\mu }^{1} &=&f_{1}(q^{2})\epsilon _{\mu \nu \alpha \beta }\varepsilon
^{\ast \nu }p^{\alpha }k^{\beta }-if_{2}(q^{2})\varepsilon _{\mu }^{\ast
}+if_{3}(q^{2})(\varepsilon ^{\ast }\cdot p)P_{\mu },  \label{60} \\
{\cal T}_{\mu }^{2} &=&f_{4}(q^{2})\epsilon _{\mu \nu \alpha \beta }\varepsilon
^{\ast \nu }p^{\alpha }k^{\beta }-if_{5}(q^{2})\varepsilon _{\mu }^{\ast
}+if_{6}(q^{2})(\varepsilon ^{\ast }\cdot p)P_{\mu }+if_{7}(q^{2})(\varepsilon ^{\ast }\cdot p)P_{\mu },   \label{61}\\
{\cal T}&=&2if_{8}(q^{2})(\varepsilon ^{\ast }\cdot p),
\end{eqnarray}
with $P_{\mu}=p_{\mu}+k_{\mu}$.

The auxiliary functions $(f's)$ which contains both long distance (form
factors) and short distance (Wilson coefficients) effects and these can be
written as
\begin{align}
f_{1}(q^{2}) =&4\widetilde{C}_{7}^{eff}\frac{m_{b}+m_{s}}{q^{2}}T_{1}(q^{2})+\widetilde{C}_{9}^{eff}\frac{2A_{V}(q^{2})}{M_{B}+M_{K^{\ast }}},  \label{621a}
\\
f_{2}(q^{2}) =&2\widetilde{C}_{7}^{eff}\frac{m_{b}-m_{s}}{q^{2}}T_{2}(q^{2})\left(
M_{B}^{2}-M_{K^{\ast }}^{2}\right) \notag\\ & +\widetilde{C}_{9}^{eff}A_{0}(q^{2})\left(
M_{B}+M_{K^{\ast}}\right),  \label{621b}\\
f_{3}(q^{2}) = &4\widetilde{C}_{7}^{eff}\frac{m_{b}-m_{s}}{q^{2}}\left( T_{2}(q^{2})+q^{2}%
\frac{T_{3}(q^{2})}{\left( M_{B}^{2}-M_{K^{\ast }}^{2}\right) }
\right) \notag\\ &+\widetilde{C}_{9}^{eff}\frac{A_{+}(q^{2})}{M_{B}+M_{K^{\ast }}},
\label{621c}\\
f_{4}(q^{2}) =&\widetilde{C}_{10}\frac{2A_{V}(q^{2})}{M_{B}+M_{K^{\ast }}},  \label{621d}
\\
f_{5}(q^{2}) =&2\widetilde{C}_{10}A_{0}(q^{2})\left( M_{B}+M_{K^{\ast }}\right),
\label{621e}\end{align}
\begin{align}
f_{6}(q^{2}) =&2\widetilde{C}_{10}\frac{A_{+}(q^{2})}{M_{B}+M_{K^{\ast }}},
\label{621f} \\
f_{7}(q^{2})=&4\widetilde{C}_{10}\frac{A_{-}(q^{2})}{M_{B}+M_{K^{\ast }}}+C_{Q2}\frac{M_{K^{\ast }}}{m(m_{b}+m_{s})} \tilde{A}_{0}(q^{2}), \label{621g}\\
f_{8}(q^{2}) =&-C_{Q1}\frac{M_{K^{\ast }}}{(m_{b}+m_{s})} \tilde{A}_{0}(q^{2}). \label{621h}
\end{align}

\section{Phenomenological Observables for $B\rightarrow K^{\ast} \ell^{+}\ell^{-}$}\label{po}

In this section we will present the calculations of the physical observables
such as the polarized branching ratio and the various lepton polarization asymmetries $P_{L,N,T}$ 
\subsection{Polarized Branching Ratio}

In the rest frame of $B$ meson the differential decay width of $%
B\rightarrow K^{\ast }\ell^{+}\ell^{-}$ can be written as
\begin{equation}
\frac{d\Gamma (B\rightarrow K^{\ast }\ell^{+}\ell^{-})}{dq^{2}}=\frac{1%
}{\left( 2\pi \right) ^{3}}\frac{1}{32M_{B}^{3}}%
\int_{-u(q^{2})}^{+u(q^{2})}du\left\vert \mathcal{M}\right\vert ^{2},
\label{62a}
\end{equation}%
where $\mathcal{M}$ is defined in Eq. (\ref{61}) and 
\begin{eqnarray}
q^{2} &=&(p_{l^{+}}+p_{l^{-}})^{2} ,\\
u &=&\left( p-p_{l^{-}}\right) ^{2}-\left( p-p_{l^{+}}\right) ^{2}.
\end{eqnarray}%
Now the limits on $q^{2}$ and $u$ are
\begin{eqnarray}
4m^{2} &\leq &q^{2}\leq (M_{B}-M_{K^{\ast }})^{2},  \label{62d} \\
-u(q^{2}) &\leq &u\leq u(q^{2}),  \label{62e}
\end{eqnarray}%
with%
\begin{equation}
u(q^{2})=\sqrt{\lambda \left( 1-\frac{4m^{2}}{q^{2}}\right) },
\label{62f}
\end{equation}%
and%
\begin{equation*}
\lambda \equiv \lambda (M_{B}^{2},M_{K^{*}}^{2},q^{2})=M_{B}^{4}+M_{K^{\ast}}^{4}+q^{4}-2M_{B}^{2}M_{K^{\ast }}^{2}-2M_{K^{\ast}}^{2}q^{2}-2q^{2}M_{B}^{2}.
\end{equation*}%
Here $m$ corresponds to the mass of the lepton which in our case are
the $\mu$ and $\tau$.  Finally, the decay rate takes the form
\begin{eqnarray}
\frac{d\Gamma}{dq^{2}} &=&\frac{G_{F}^{2}\left\vert
V_{tb}V_{ts}^{\ast }\right\vert ^{2}\alpha ^{2}}{2^{11}\pi
^{5}3M_{B}^{3}M_{K^{\ast }}^{2}q^{2}}u(q^{2})\times \mathcal{B}\left(
q^{2}\right)  \label{62i} \\
\end{eqnarray}%
The function $u(q^2)$ is defined in Eq. (\ref{62f}) and  $\mathcal{B}(q^2)$ is
\begin{align}
\mathcal{B}(q^{2}) &=8M_{K^{\ast }}^{2}q^{2}\lambda
\bigg\{(2m^{2}+q^{2})\left\vert f_{1}(q^{2})\right\vert ^{2}
-(4m^{2}-q^{2})\left\vert f_{4}(q^{2})\right\vert ^{2}\bigg\}  +4M_{K^{\ast }}^{2}q^{2}\bigg\{(2m^{2}+q^{2})\notag \\
&\times\left( 3\left\vert f_{2}(q^{2})\right\vert ^{2}-\lambda \left\vert f_{3}(q^{2})\right\vert
^{2}\right) -(4m^{2}-q^{2})\left( 3\left\vert f_{5}(q^{2})\right\vert ^{2}-\lambda
\left\vert f_{6}(q^{2})\right\vert ^{2}\right) \bigg\}\notag \\
&+\lambda (2m^{2}+q^{2})\left\vert
f_{2}(q^{2})+\left(M_{B}^{2}-M_{D_{s}^{\ast
}}^{2}-q^{2}\right)f_{3}(q^{2})\right\vert ^{2}+ 24m^{2}M_{D_{s}^{\ast
}}^{2}\lambda\left\vert f_{7}(q^{2})\right\vert^{2} \notag \\
&-(4m^{2}-q^{2})\left\vert f_{5}(q^{2})+\left(M_{B}^{2}-M_{D_{s}^{\ast
}}^{2}-q^{2}\right)f_{6}(q^{2})\right\vert ^{2}  +\left(q^{2}-4 m^2\right) \lambda\left\vert f_{8}(q^{2})\right\vert ^{2}  \notag \\
& -12 m^{2}q^{2}\left[\Re(f_{5}f_{7}^{\ast })-\Re(f_{6}f_{7}^{\ast })\right].\label{63b}
\end{align}

In case, when the final state meson, i.e., $K^{\ast}$ is polarized, the total decay rate for the decay $B\rightarrow
K^{\ast }\ell^{+}\ell^{-}$ can be written in terms of longitudinal $\Gamma_{L}$ and transverse component $\Gamma_T$ as \cite{AAIJA}:\
\begin{eqnarray}
\frac{d\Gamma_L}{dq^{2}} &=&\frac{G_{F}^{2}\left\vert
V_{tb}V_{ts}^{\ast }\right\vert ^{2}\alpha ^{2}}{2^{11}\pi
^{5}M_{B}^{3}}u(q^{2})\times \frac{1}{(3M_K^{*}q^2)} \mathcal{A}_L\left(
q^{2}\right) , \label{62j} \\
\frac{d\Gamma_{T}}{dq^{2}} &=&\frac{G_{F}^{2}\left\vert
V_{tb}V_{ts}^{\ast }\right\vert ^{2}\alpha ^{2}}{2^{11}\pi
^{5}M_{B}^{3}}u(q^{2})\times \frac{4}{3} \mathcal{A}_{T}\left(
q^{2}\right),  \label{62k}
\end{eqnarray}
where the functions $\mathcal{A}_L$ and $\mathcal{A}_{\pm}$ are given in the appendix.

Finally, the polarized branching ratio can be written as
\begin{equation}
\mathcal{B}_{L,T} = \frac{\int_{q^2_{min}}^{q^2_{max}}\frac{d\Gamma_{L,T}}{dq^2}dq^2}{\Gamma_{tot}} \label{pol-branching},
\end{equation}
where $\Gamma_{tot}$ is the total decay width of the $B$ decay and its value is obtained from \cite{pdg}.


\subsection{Lepton Polarization Asymmetries}

In the rest frame of the lepton $\ell^{-}$, the unit vectors along
longitudinal, normal and transversal component of the $\ell^{-}$ can be defined
as \cite{jam,Aliev,22,23,aa,aa1,AAIJA}:
\begin{subequations}
\begin{eqnarray}
s_{L}^{-\mu } &=&(0,\vec{e}^{-}_{L})=\left( 0,\frac{\vec{p}_{-}}{\left| \vec{p}%
_{-}\right| }\right) , \label{p-vectorsa} \\
s_{N}^{-\mu } &=&(0,\vec{e}^{-}_{N})=\left( 0,\frac{\vec{k} \times
\vec{p}_{-}}{\left| \vec{k}\times \vec{p}_{-}\right| }\right) ,
\label{p-vectorsb} \\
s_{T}^{-\mu } &=&(0,\vec{e}^{-}_{T})=\left( 0,\vec{e}_{N}\times \vec{e}%
_{L}\right) ,  \label{p-vectorsc}
\end{eqnarray}
\end{subequations}
where $\vec{p}_{-}$ and $\vec{k}$ are the three-momenta of the
lepton $\ell^{-}$ and $K^{*}$ meson, respectively, in the center mass
(c.m.) frame of $\ell^{+}\ell^{-}$ system. Lorentz transformation is used to boost
the longitudinal component of the lepton polarization to the c.m. frame of the
lepton pair as
\begin{equation}
\left( s_{L}^{-\mu }\right) _{CM}=\left( \frac{|\vec{p}_{-}|}{m},\frac{%
E\vec{p}_{-}}{m\left| \vec{p}_{-}\right| }\right),
\label{bossted component}
\end{equation}
where $E$ and $m$ are the energy and mass of the lepton. The normal
and transverse components remain unchanged under the Lorentz boost. The
longitudinal ($P_{L}$), normal ($P_{N}$) and transverse ($P_{T}$)
polarizations of lepton can be defined as:
\begin{equation}
P_{i}^{(\mp )}(q^{2})=\frac{\frac{d\Gamma }{dq^{2}}(\vec{\xi}^{\mp }=\vec{e}%
^{\mp })-\frac{d\Gamma }{dq^{2}}(\vec{\xi}^{\mp }=-\vec{e}^{\mp })}{\frac{%
d\Gamma }{dq^{2}}(\vec{\xi}^{\mp }=\vec{e}^{\mp })+\frac{d\Gamma }{dq^{2}}(%
\vec{\xi}^{\mp }=-\vec{e}^{\mp })},  \label{polarization-defination}
\end{equation}%
where $i=L,\;N,\;T$ and $\vec{\xi}^{\mp }$ is the spin direction along the
leptons $\ell^{\mp }$. The differential decay rate for polarized lepton $\ell^{\mp
}$ in $B\rightarrow D_{s}^{*}\ell^{+}\ell^{-}$ decay along any spin
direction $\vec{\xi}^{\mp }$ is related to the unpolarized decay rate (\ref{62i}) with the following relation
\begin{equation}
\frac{d\Gamma (\vec{\xi}^{\mp })}{dq^{2}}=\frac{1}{2}\left( \frac{d\Gamma }{%
dq^{2}}\right) \left[1+(P_{L}^{\mp }\vec{e}_{L}^{\mp }+P_{N}^{\mp }\vec{e}%
_{N}^{\mp }+P_{T}^{\mp }\vec{e}_{T}^{\mp })\cdot \vec{\xi}^{\mp }\right].
\label{polarized-decay}
\end{equation}%
The expressions of the longitudinal, normal and transverse lepton
polarizations can be written as
\begin{align}
P_{L}(q^2) \propto &\frac{4\lambda}{3M_{K^{\ast }}^{2}}\sqrt{\frac{q^2-4m^{2}}{q^{2}}}\times  \bigg\{2\Re(f_{2}f_{5}^{\ast})+\lambda\Re(f_{3}f_{6}^{\ast})+4\sqrt{q^{2}}\Re(f_{1}f_{4}^{\ast})\left(1+\frac{12q^2M_{K^{\ast}}^{2}}{\lambda}\right)\notag\\
& +\left(-M_{B}^{2}+M_{K^{\ast }}^{2}+q^{2}\right)\left[\Re(f_{3}f_{5}^{\ast})+\Re(f_{2}f_{6}^{\ast})\right]\notag\\
&+ \frac{3}{2}m\left[\Re(f_{5}f_{8}^{\ast})+\Re(f_{6}f_{8}^{\ast})\left(-M_{B}^{2}+M_{K^{\ast }}^{2}\right)-\Re(f_{7}f_{8}^{\ast})\right]\bigg\}, \label{long-polarization}
\end{align}
\begin{align}
P_{N}(q^2) \propto & \frac{m\pi}{M_{K^{\ast }}^{2}}\sqrt{\frac{\lambda}{q^{2}}}\times \bigg\{-\lambda q^{2}\Re(f_{3}f_{7}^{\ast})+\lambda(M_{B}^{2}-M_{K^{\ast }}^{2})\Re(f_{3}f_{6}^{\ast})-\lambda\Re(f_{3}f_{5}^{\ast})\notag\\
&+\left(-M_{B}^{2}+M_{K^{\ast }}^{2}+q^{2}\right)\left[q^{2}\Re(f_{2}f_{7}^{\ast})+(M_{B}^{2}-M_{K^{\ast }}^{2})\Re(f_{2}f_{5}^{\ast})+(q^{2}-4m^2)\Re(f_{5}f_{8}^{\ast})\right]\notag\\
& - 8q^{2}M_{K^{\ast }}^{2}\Re(f_{1}f_{2}^{\ast}) +\sqrt{\lambda }(q^{2}-4m^2)\Re(f_{6}f_{8}^{\ast})\bigg\},   \label{norm-polarization}\\
P_{T}\left( q^{2}\right)  \propto & i\frac{m\pi \sqrt{ \left( q^{2}-\frac{4m^{2}}{q^{2}}\right) \lambda }}{M_{K^{\ast }}^{2}}\bigg\{M_{K^{\ast}}\left[4\Im(f_{2}f_{4}^{\ast})+4\Im(f_{1}f_{5}^{\ast})+3\Im(f_{5}f_{6}^{\ast}) \right]\notag\\
&-\lambda\Im(f_{6}f_{7}^{\ast})+\left(-M_{B}^{2}+M_{K^{\ast }}^{2}+q^{2}\right)\bigg[\Im(f_{7}f_{5}^{\ast})+\Im(f_{2}f_{8}^{\ast})\bigg]-q^{2}\Im(f_{5}f_{6}^{\ast})\bigg\}.
\label{Transverse-polarization}
\end{align}%
Here we have dropped out the constant factors which, however, are understood.
The average value of different lepton polarization asymmetries can be written as:
\begin{equation}
\langle \mathcal{P}_i \rangle = \frac{\int_{s_{min}}^{s_r}\mathcal{P}_i \frac{d\mathcal{B}}{ds}ds}{\int_{s_{min}}^{s_r}\frac{d\mathcal{B}}{ds}ds}
\end{equation}
where the subscript $i$ can be $L, T$ and $\mathcal{B}$ is the branching ratio. The upper limit of integration, i.e., $s_r$ corresponding to the value of momentum transfer below the resonance region.

\section{Numerical Results and Discussion}\label{num}

This section is dedicated to the numerical results of the polarized branching ratio $\mathcal{BR}_{L, T}$ and different lepton polarizations asymmetries $P_{L,N,T}$ for the $B\rightarrow
K^{\ast }\ell^{+}\ell^{-}$ decays with $\ell=\mu,\tau$. At first the numerical values of input parameters which are used in our numerical
calculations are given in Table \ref{input} \cite{pdg}:
\begin{table}[ht]
\centering
\begin{tabular}{l}
\hline\hline
$M_{B}=6.277$ GeV, $m_{K^{\ast}}=2.1123$ GeV, $m_{b}=4.28$ GeV, $m_{\mu}=0.105$ GeV\\
$m_{\tau}=1.77$ GeV, $|V_{tb}V_{ts}^{\ast}|=4.1\times
10^{-2}$, $\tau_{B}=0.453\times 10^{-12}$ sec\\
$G_{F}=1.17\times 10^{-5}$ GeV$^{-2}$, $\alpha^{-1}=137$, $f_B$ = 0.35 GeV\\
\hline\hline
\end{tabular}
\caption{Values of input parameters used in our numerical analysis \cite{pdg}.}\label{input}
\end{table}

\begin{table*}[ht]
\centering \caption{The Wilson coefficients $C_{i}^{\mu}$ at the
scale $\mu\sim m_{b}$ in the SM .}
\begin{tabular}{cccccccccc}
\hline\hline
$C_{1}$&$C_{2}$&$C_{3}$&$C_{4}$&$C_{5}$&$C_{6}$&$C_{7}$&$C_{9}$&$C_{10}$
\\ \hline
 \ \  1.107 \ \  & \ \  -0.248 \ \  & \ \  -0.011 \ \  & \ \  -0.026 \ \  & \ \  -0.007 \ \  & \ \  -0.031 \ \  & \ \  -0.313 \ \  & \ \  4.344 \ \  & \ \  -4.669 \ \  \\
\hline\hline
\end{tabular}
\label{wc table}
\end{table*}
Of course to perform the numerical analysis, another important ingredient is the form factors. The values of the form factors used in the upcoming analysis are
the ones calculated using the QCD sum rules and these are summarized in Table I.

In the model under discussion, i.e., the THDMs, the free parameters in these models are the masses of charged Higgs boson $m_{H^{\pm}}$, the coefficients $\lambda_{tt}$, $\lambda_{bb}$ and the ratio of the vacuum expectation values of the two Higgs doubles, i.e. $tan\beta$. The coefficients $\lambda_{tt}$ and $\lambda_{bb}$ for the version I and II of the THDM are:
\begin{eqnarray}
\lambda_{tt} = \cot\beta,\hskip 1em  \lambda_{bb} = - \cot\beta, \hskip 1em \textit{for model I}, \notag \\
\lambda_{tt} = \cot\beta, \hskip 1em \lambda_{bb} =  +\tan\beta, \hskip 1em \textit{for model II}. \label{lambdas}
\end{eqnarray}
and for version III of THDM, these coefficients are complex, i.e.,
\begin{equation}
\lambda_{tt}\lambda_{bb} \equiv |\lambda_{tt}\lambda_{bb} | e^{i \delta},
\end{equation}
where $\delta$ is a single $CP$ phase of the vacuum in this version.

To constraint the mass of charged Higgs boson and $\tan \beta$ one can use the experimental results of the branching ratio of $b\to s \gamma$ and $B \to D \ell \nu_{\ell}$ decays as well as $B - \bar{B}$ and $K - \bar{K}$ mixing in the literature \cite{constraints}. In addition, the parameters 
$|\lambda_{tt}|$,  $|\lambda_{bb}|$ and the phase $\delta$ are restricted by the experimental results of the electric dipole moments of neutron, $B - \bar{B}$ mixing, 
$\rho_0$, $R_b$ and $Br(b \to s \gamma)$ \cite{constraint1, constraint2, constraint3, constraint4}. The value of $\lambda_{tt}\lambda_{bb}$ is constrained to be 1 and the $\delta$ is restricted in the range $60^{\circ} - 90^{\circ}$ by using the experimental limits on electric dipole moment of neutron and $Br(b \to s \gamma)$, plus the constraint on $M_{H^{+}}$ from the LEP II. Using the constraints from the $B - \bar{B}$ mixing as well as from $R_b$, the analysis of various lepton polarization asymmetries in $B \to K^{\ast} \ell^{+}\ell^{-}$ has been done in the following parametric space in model III \cite{Falahati}:
\begin{eqnarray}
\textit{Case A}: \hskip 1em |\lambda_{tt}| = 0.03,\hskip 1em  |\lambda_{bb}| = 100, \hskip 1em \notag \\
\textit{Case B}: \hskip 1em |\lambda_{tt}| = 0.15,\hskip 1em  |\lambda_{bb}| = 50, \hskip 1em \label{lamdavalues} \\
\textit{Case C}: \hskip 1em |\lambda_{tt}| = 0.3,\hskip 1em  |\lambda_{bb}| = 30. \hskip 1em \notag
\end{eqnarray}
Where $\delta =\pi/2$ and the values of masses of Higgs particles are summarized in Table 4:
\begin{table}[ht]
\centering
\begin{tabular}{lcccc}
\hline\hline
Masses (GeV)  & $m_{A^0}$&  $m_{h^0}$& $m_{H^0}$& $m_{H^\pm}$ \\
\hline
Set I &  $125$& $125$&  $160$& $200$\\
Set II & $125$&  $125$& $160$& $160$\\
Set III &  $125$& $125$&  $125$& $200$\\
Set IV & $125$&  $125$& $125$& $160$\\ 
\hline\hline
\end{tabular}
\caption{Values of the masses of the Higgs particles.}\label{susy2}
\end{table}
It is an established fact that in THDM of type II the charged Higgs contribution to $B \to \tau \nu$ interferes necessarily destructive with the SM contributions \cite{wu}. The enhancement of $Br(B \to \tau \nu)$ is possible if the absolute value of the contribution of the charged Higgs boson is two times the SM one, which then conflict  with the $B \to D \tau \nu$ decay. Furthermore, this version of THDM failed to explain the observed discrepancy of $2.2\sigma$ in $R(D)$ and $2.7\sigma$ in $R(D^*)$ compared to their SM value. In order to cure this situation, a detailed discussion on the model III has been done in Ref. \cite{Greub}. The purpose of present study is not to put the precise bounds on the parameters of versions of THDM but is to check the profile of different physical observables, e.g., the polarized branching ratio and the  lepton polarization asymmetries in $B \to K^* \ell^{+} \ell^{-}$ decays in THDM of type III.  Here, we would like to mention that similar study exist in the literature \cite{Falahati:2015hxa} but the choice of observables is different as well as the important contribution form the charm-loop in Wilson coefficient $C_9^{eff}$ is ignored.

\begin{figure}[ht]
\begin{tabular}{cc}
\centering
\hspace{0.5cm}($\mathbf{a}$)&\hspace{1.2cm}($\mathbf{b}$)\\
\includegraphics[scale=0.55]{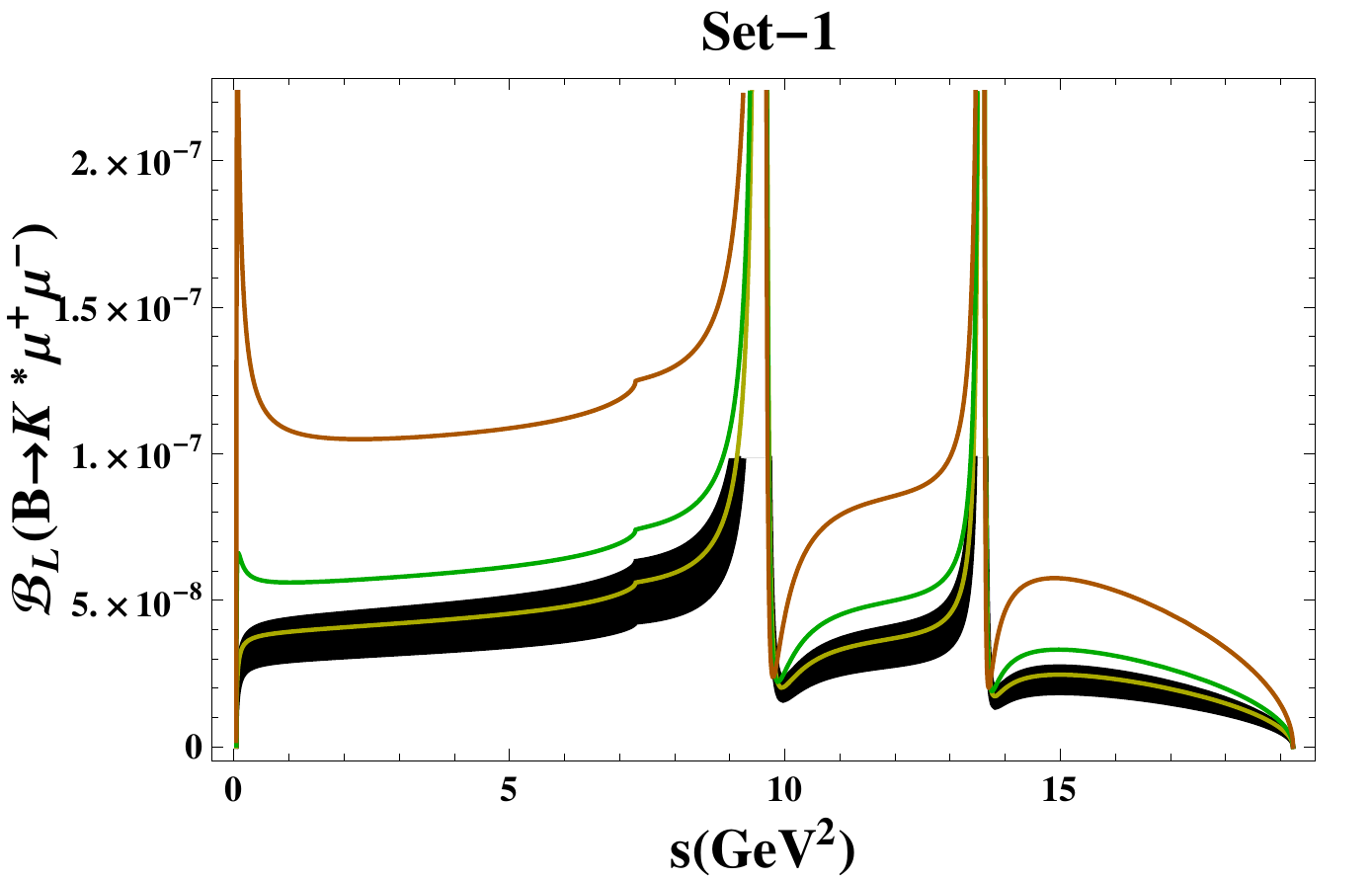}&\includegraphics[scale=0.55]{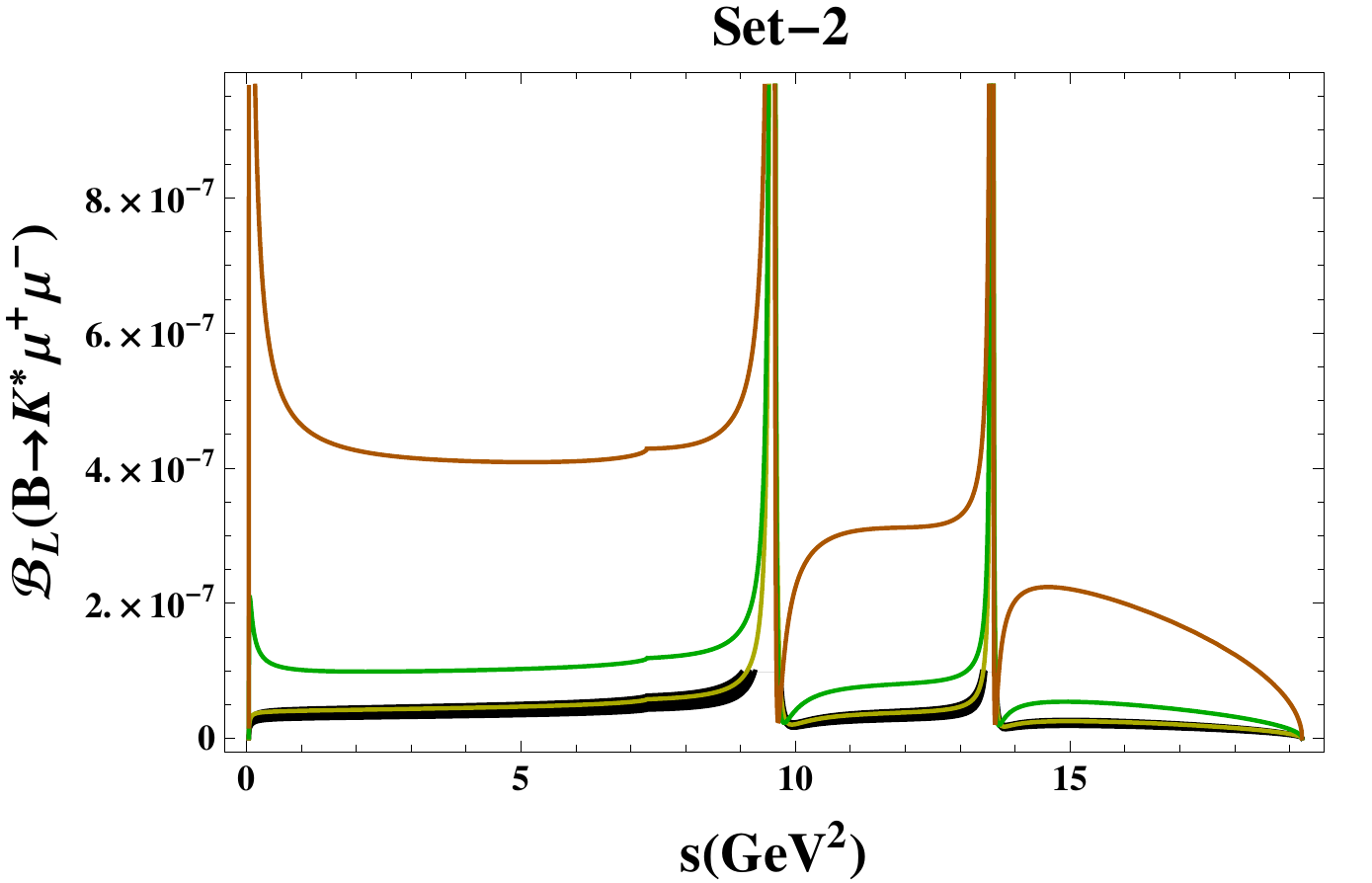}\\
\hspace{0.5cm}($\mathbf{c}$)&\hspace{1.2cm}($\mathbf{d}$)\\
\includegraphics[scale=0.55]{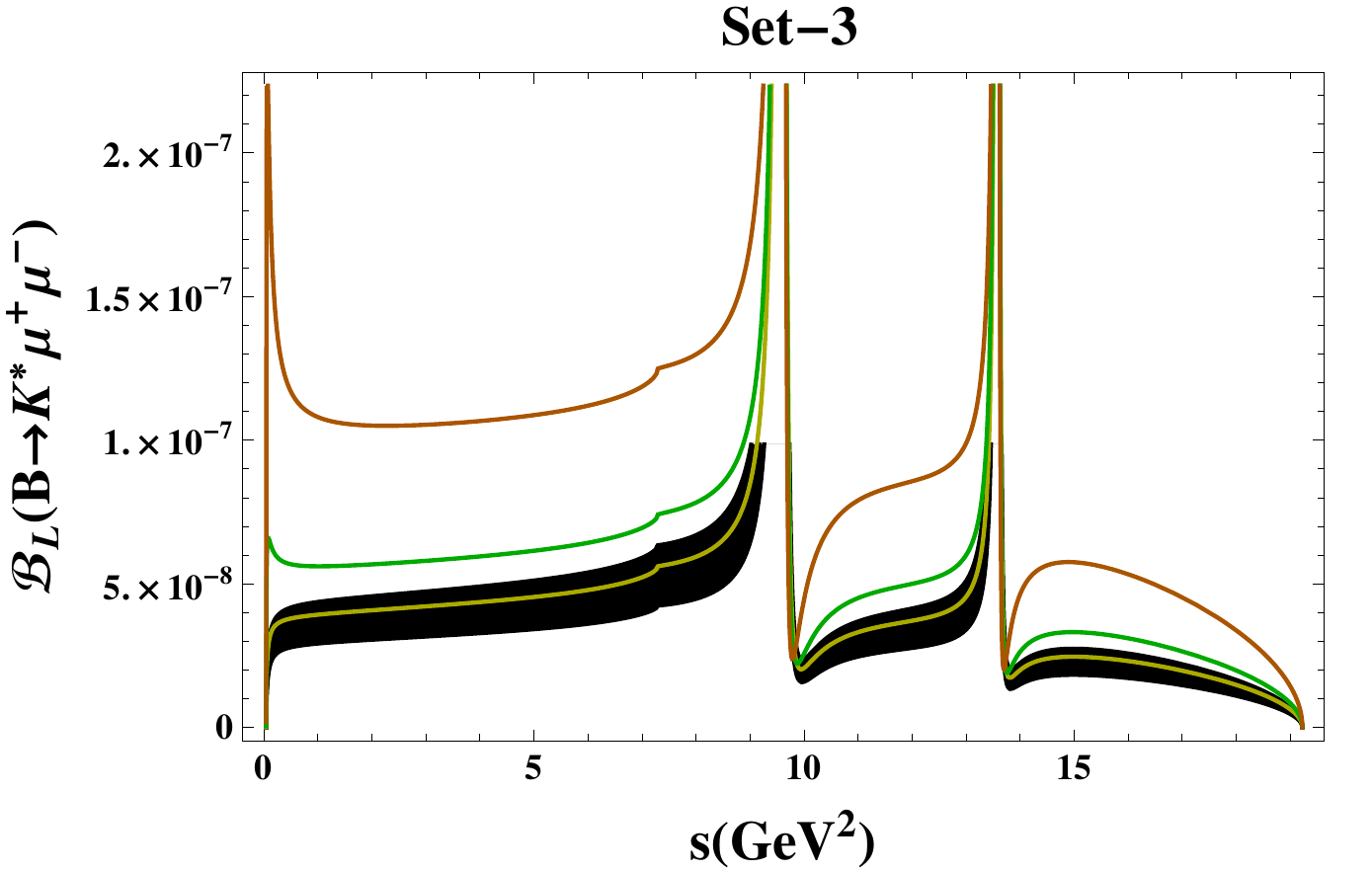}&\includegraphics[scale=0.55]{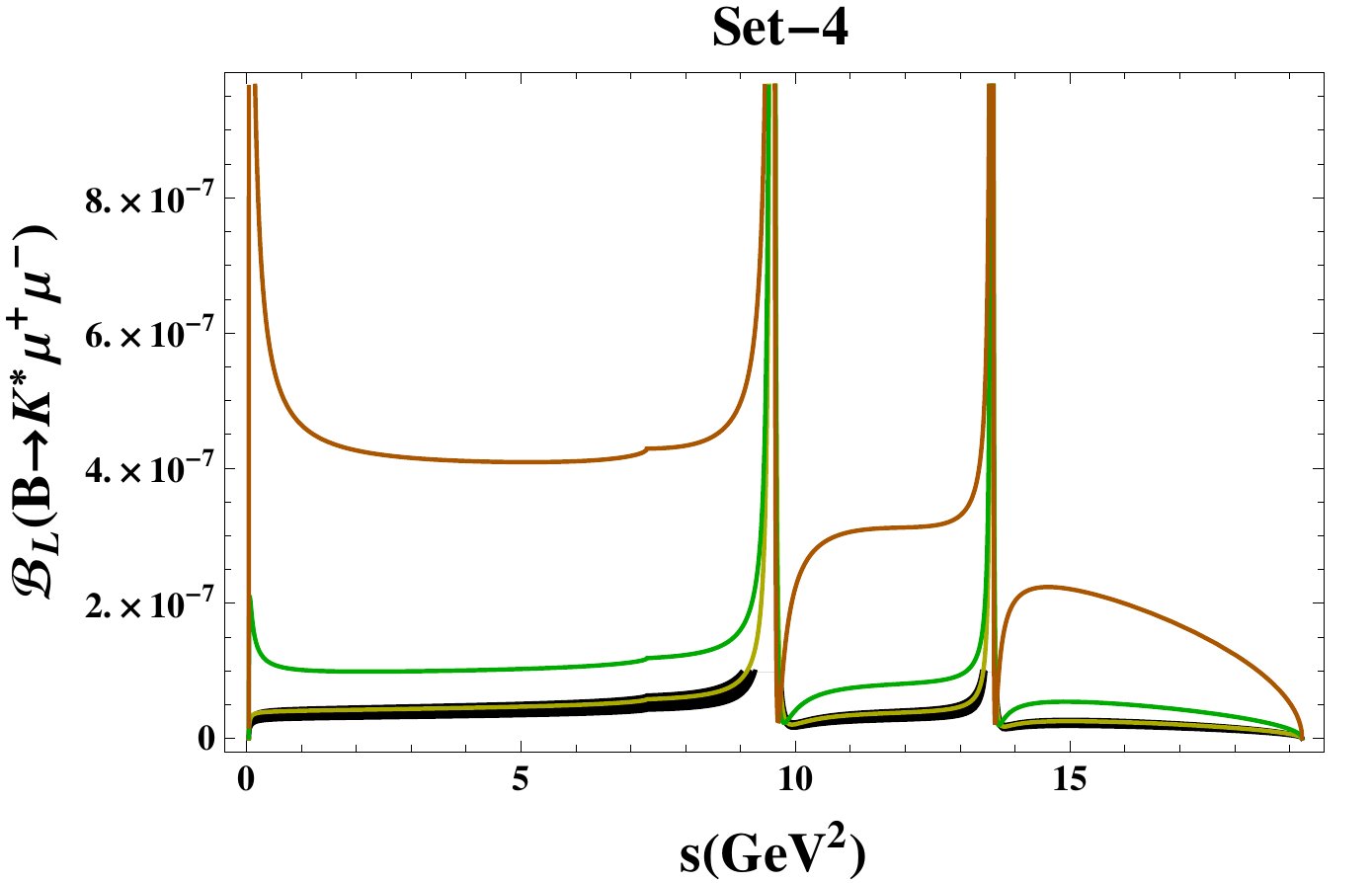}\end{tabular}
\caption{The dependence of longitudinally polarized branching ratio for $B\to K^{*}\mu^{+}\mu^{-}$ decay on $s$. The yellow, green and orange lines correspond to the \textit{Case A}, \textit{Case B} and \textit{Case C}, respectively.  In all the graphs solid band corresponds to the SM uncertainties.} \label{longbrmu}
\end{figure}

\begin{figure}[ht]
\begin{tabular}{cc}
\centering
\hspace{0.5cm}($\mathbf{a}$)&\hspace{1.2cm}($\mathbf{b}$)\\
\includegraphics[scale=0.55]{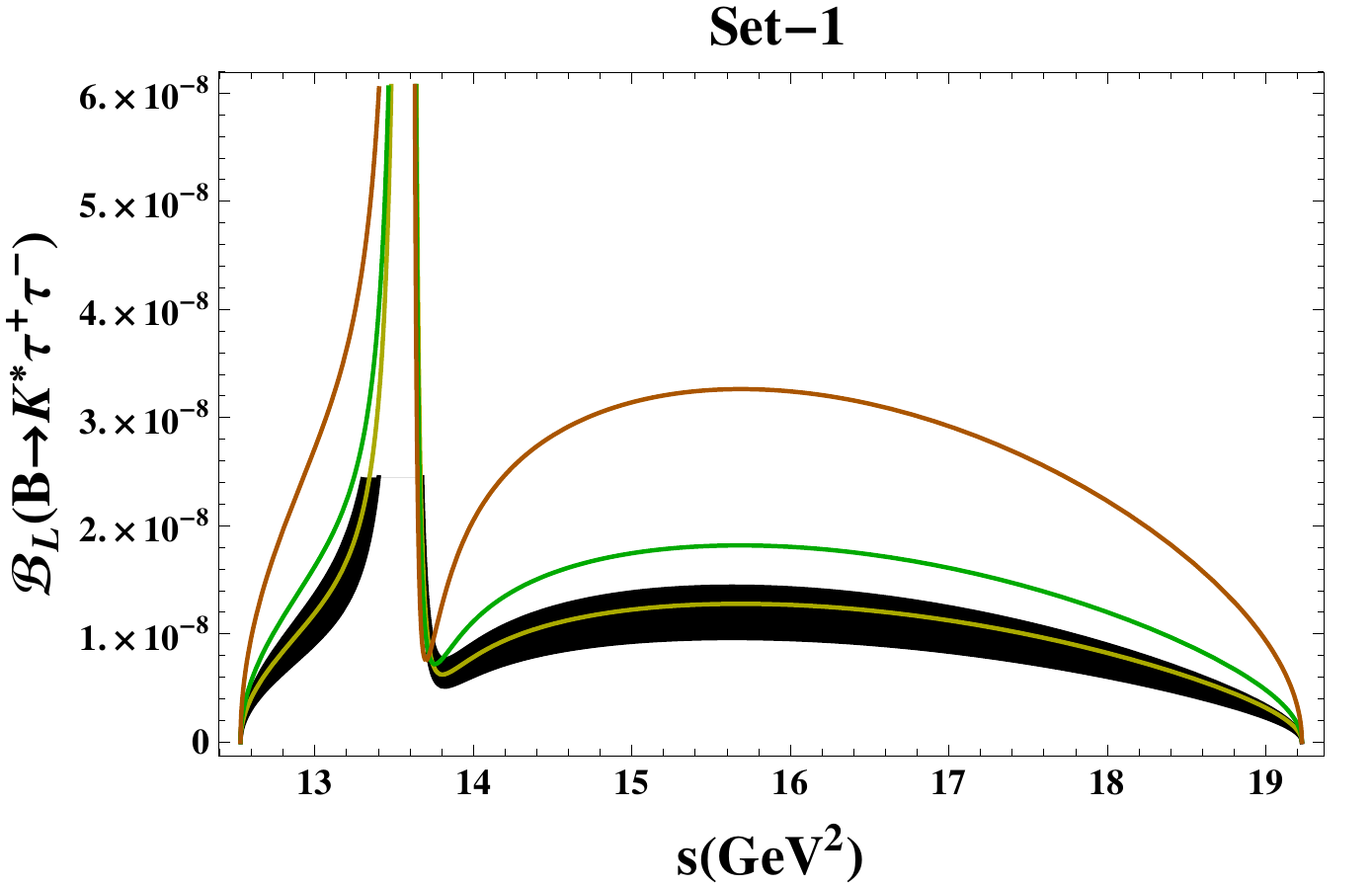}&\includegraphics[scale=0.55]{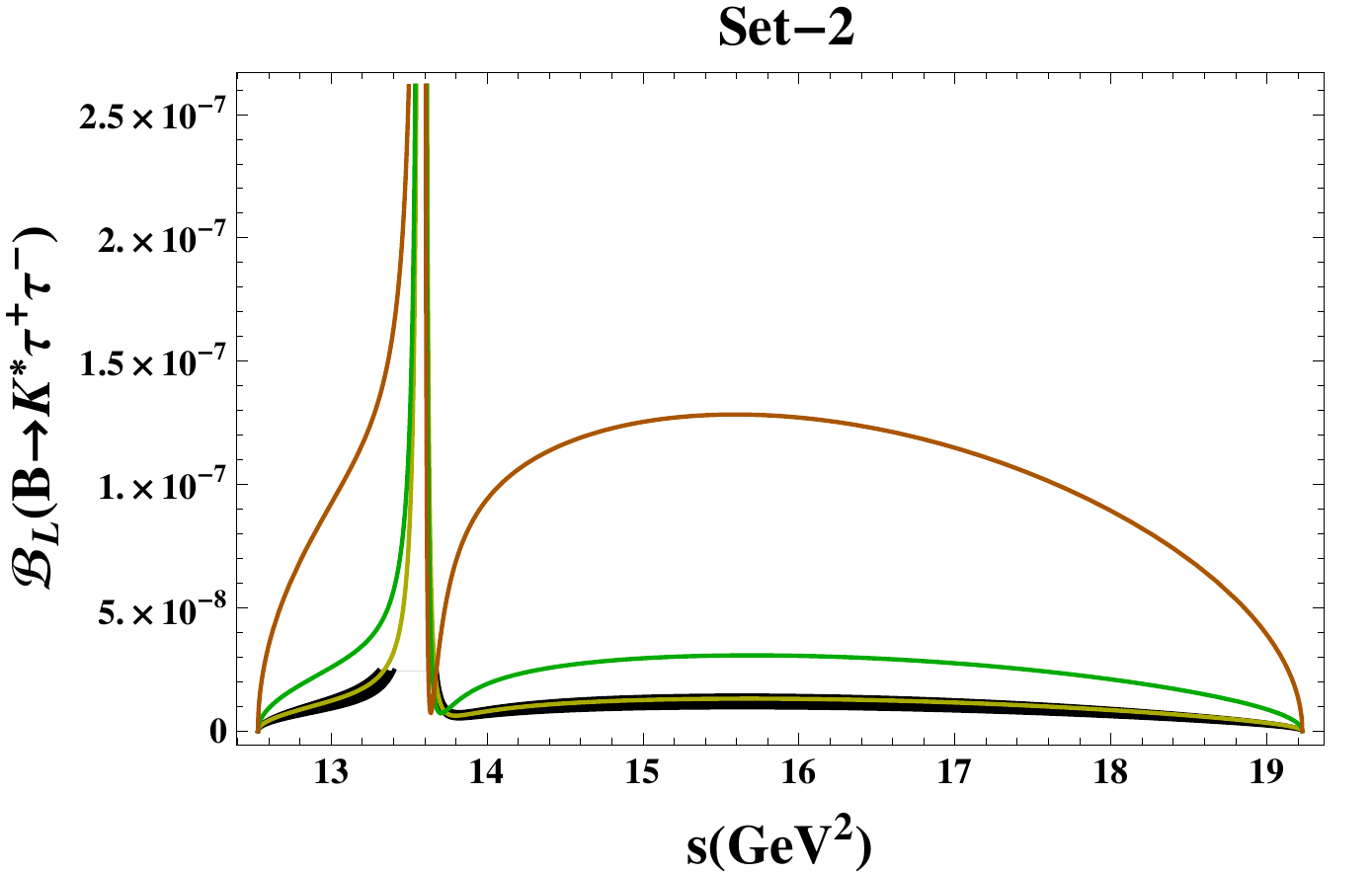}\\
\hspace{0.5cm}($\mathbf{c}$)&\hspace{1.2cm}($\mathbf{d}$)\\
\includegraphics[scale=0.55]{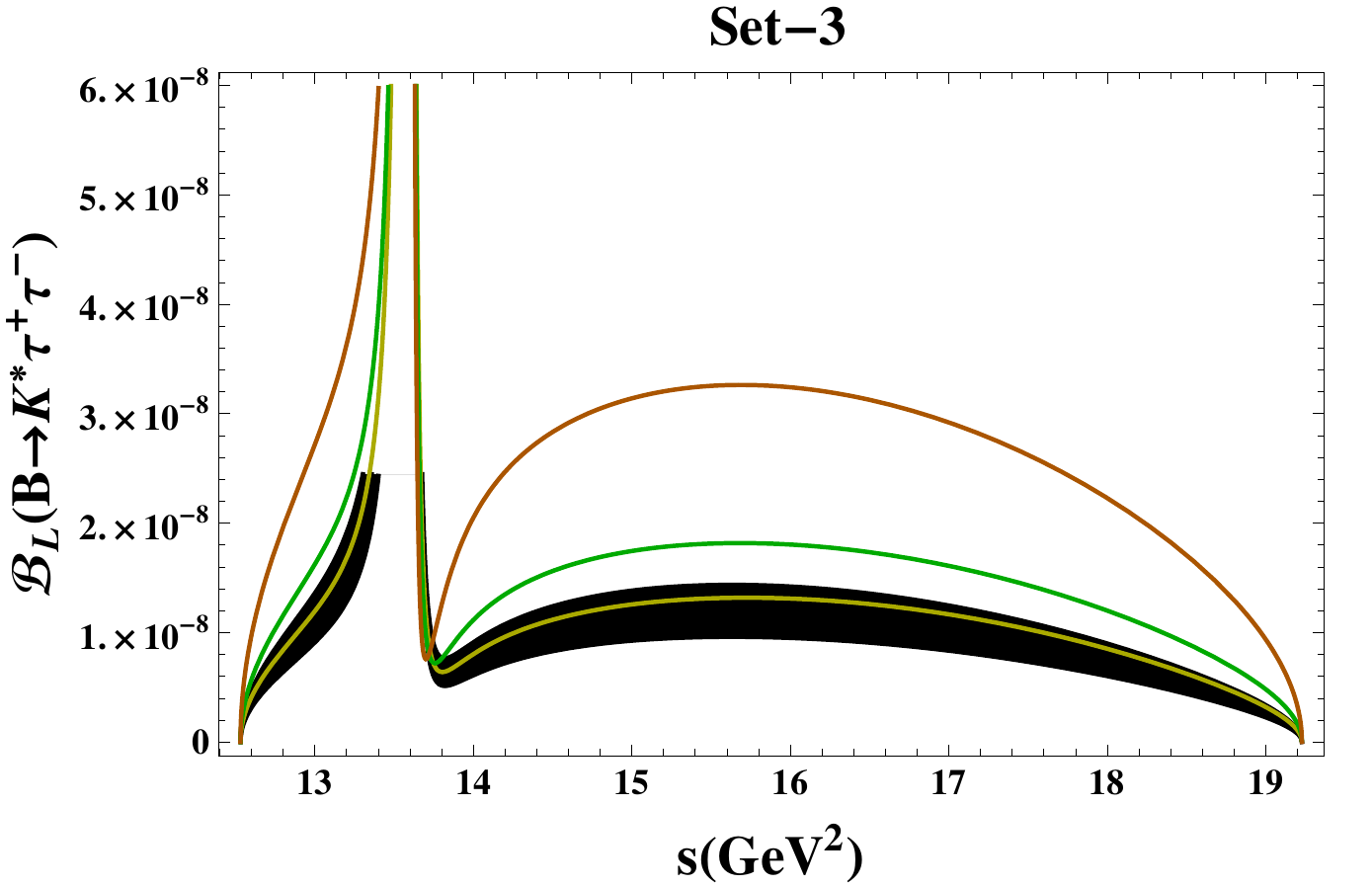}&\includegraphics[scale=0.55]{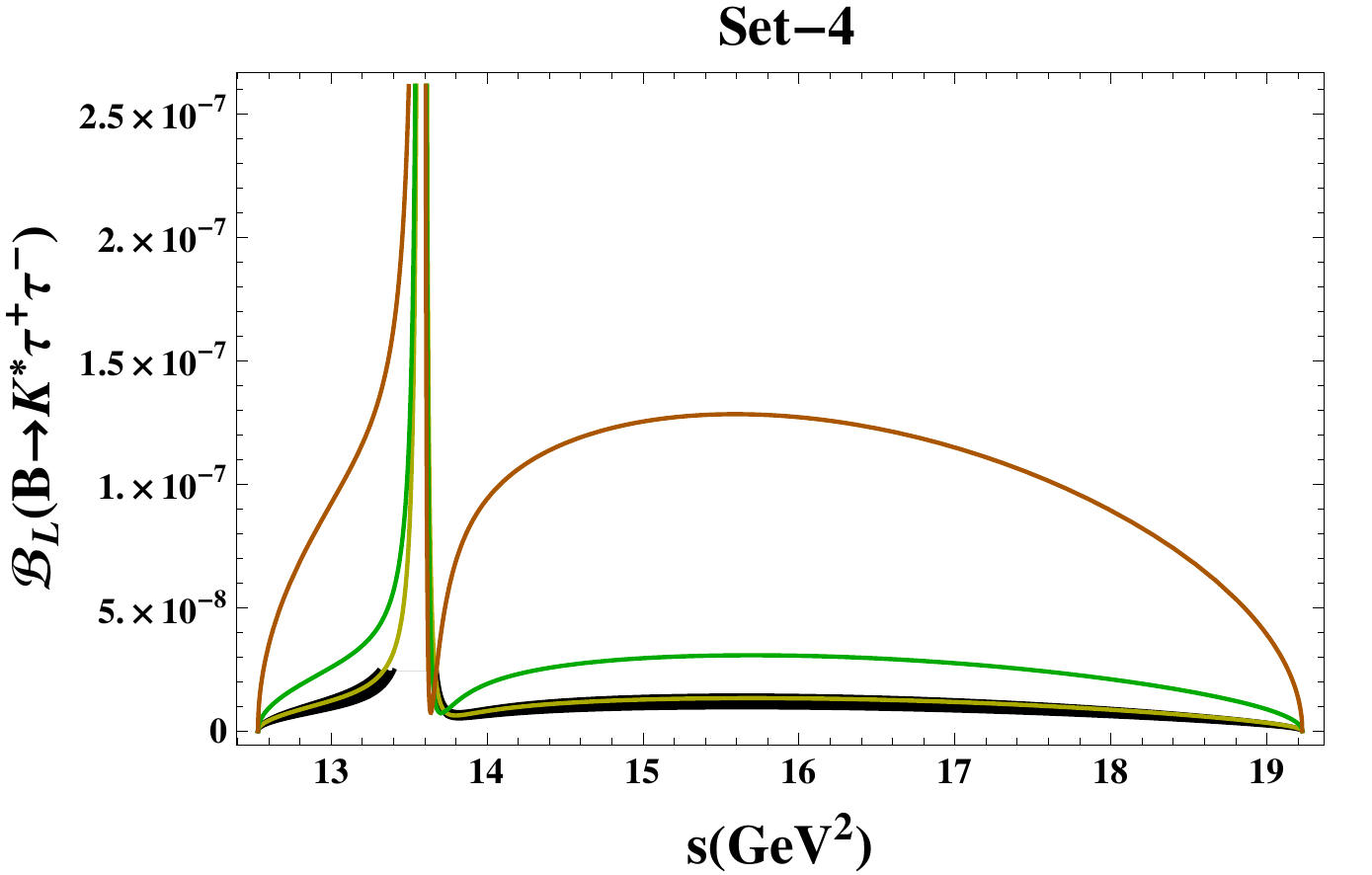}\end{tabular}
\caption{The dependence of longitudinally polarized branching ratio for $B\to K^{*}\tau^{+}\tau^{-}$ decay on $s$. The legends are same as in Fig. \ref{longbrmu}.} \label{longbrtau}
\end{figure}

\begin{figure}[ht]
\begin{tabular}{cc}
\centering
\hspace{0.5cm}($\mathbf{a}$)&\hspace{1.2cm}($\mathbf{b}$)\\
\includegraphics[scale=0.55]{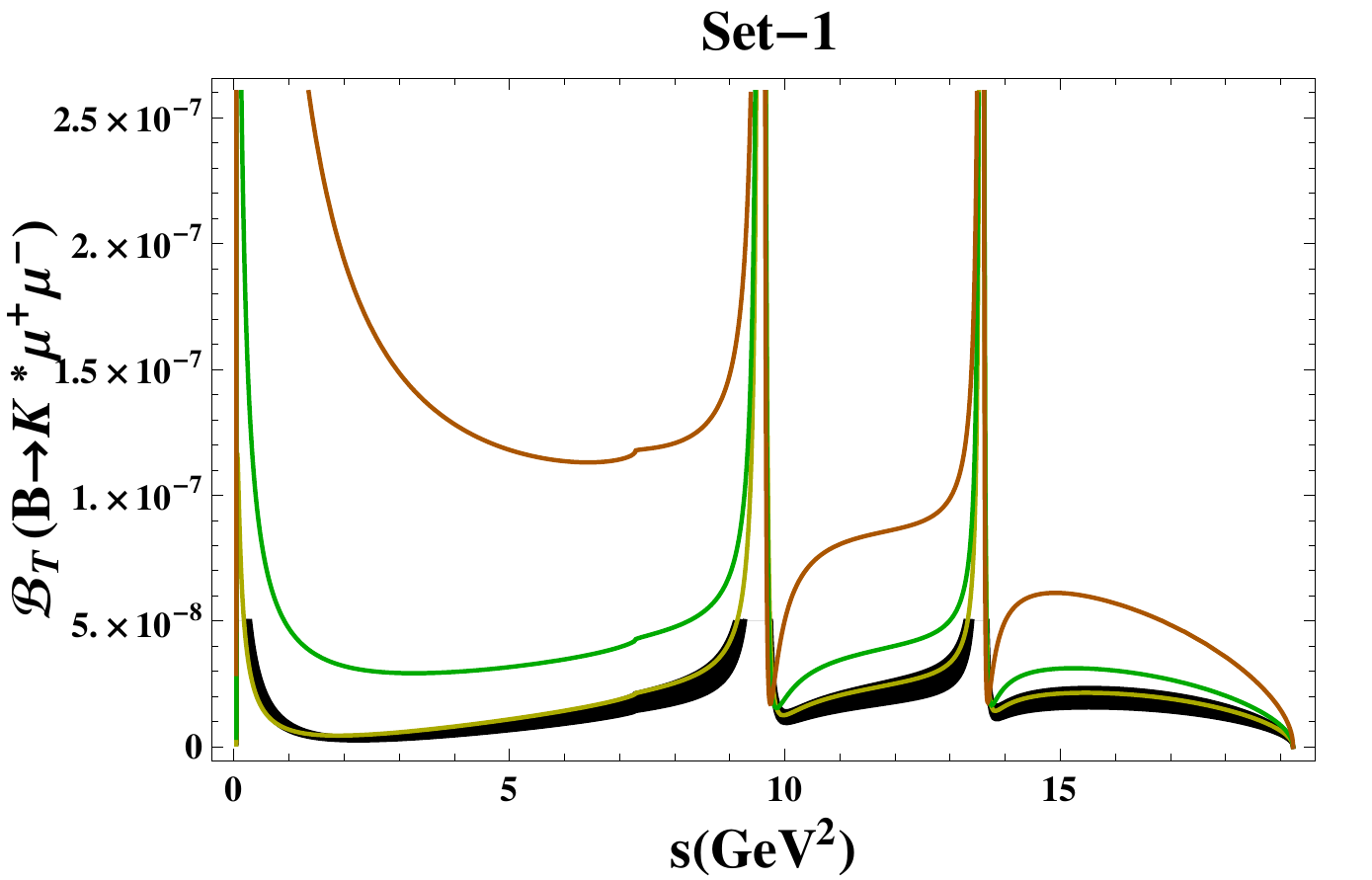}&\includegraphics[scale=0.55]{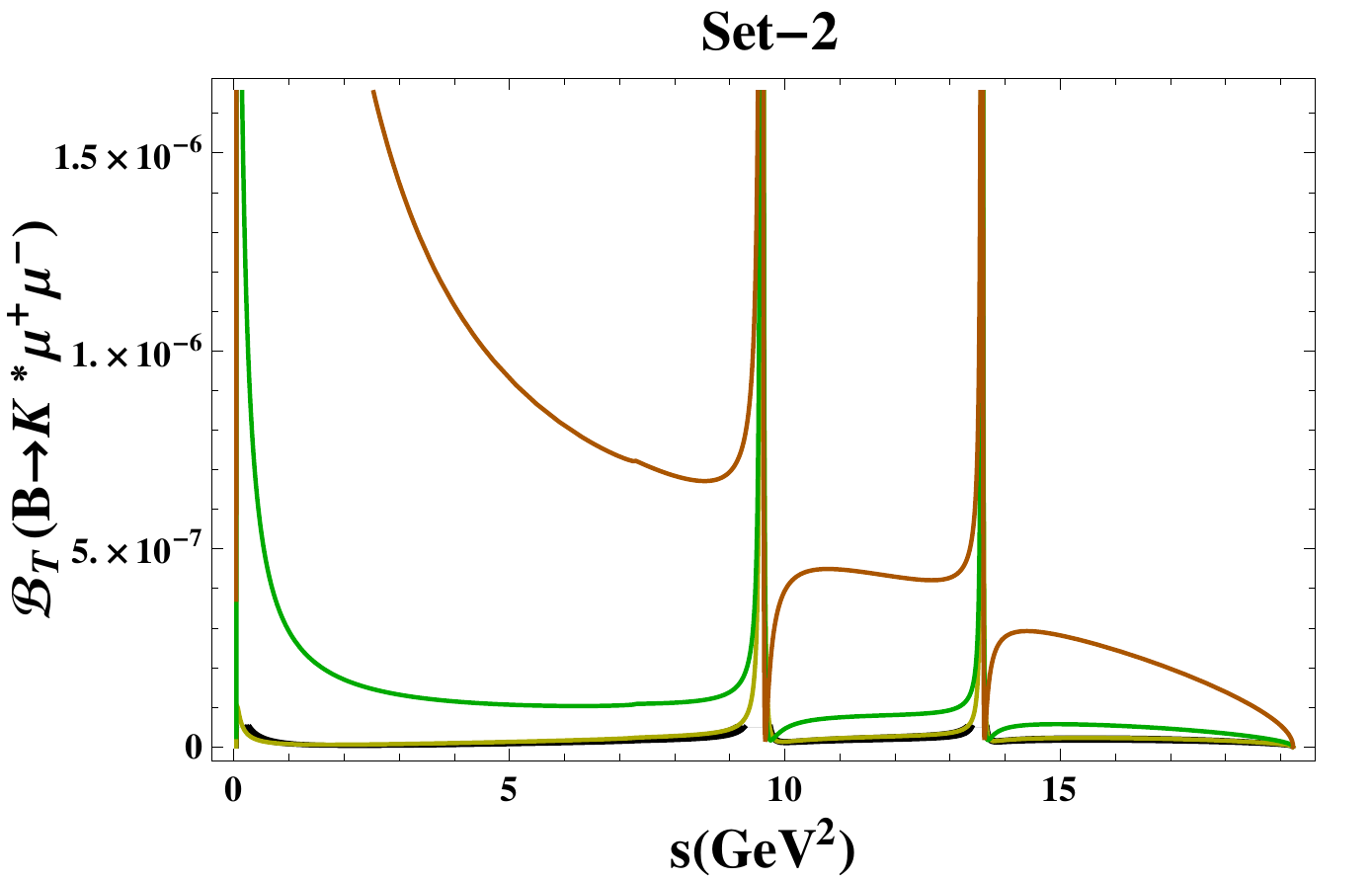}\\
\hspace{0.5cm}($\mathbf{c}$)&\hspace{1.2cm}($\mathbf{d}$)\\
\includegraphics[scale=0.55]{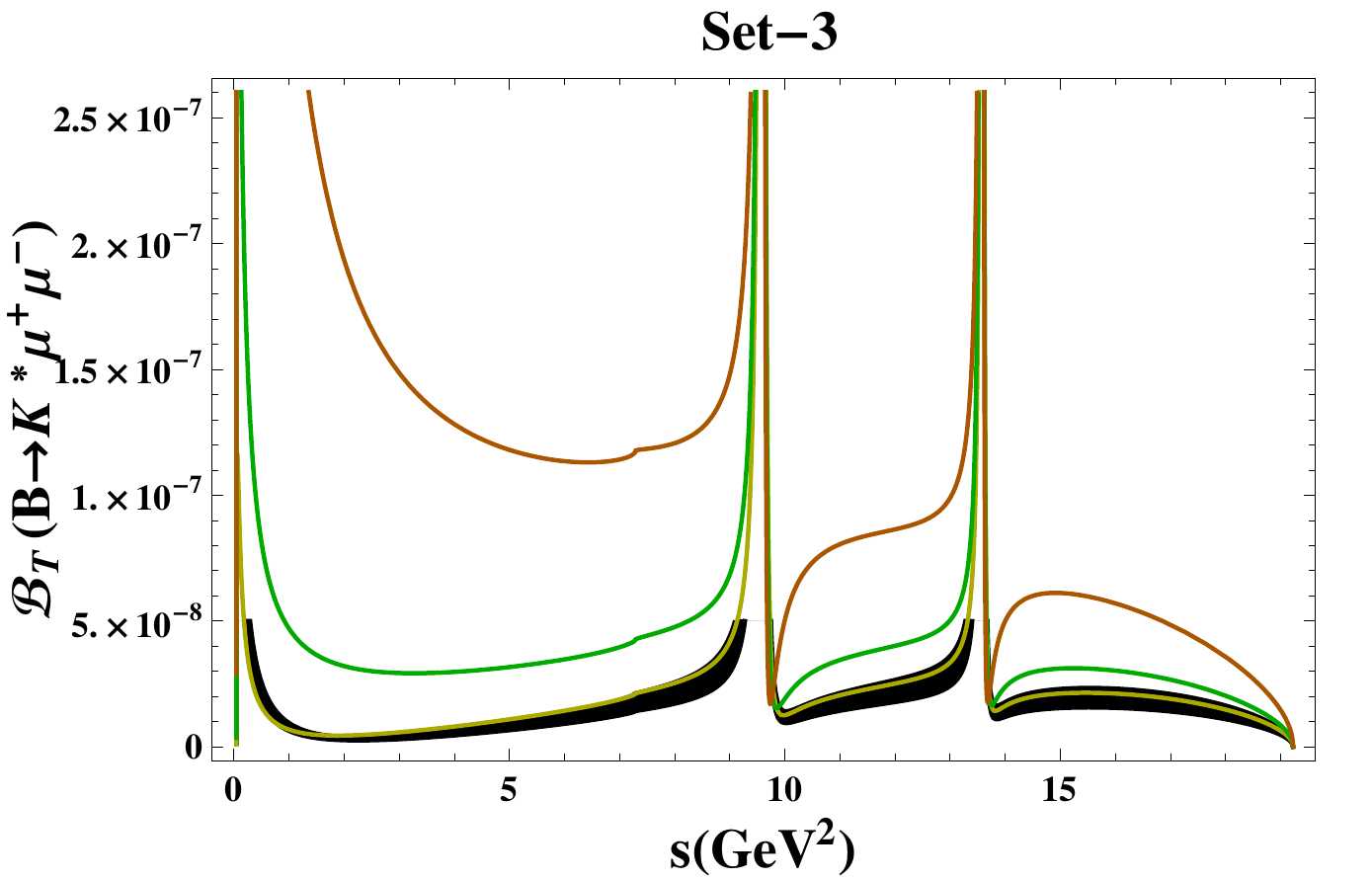}&\includegraphics[scale=0.55]{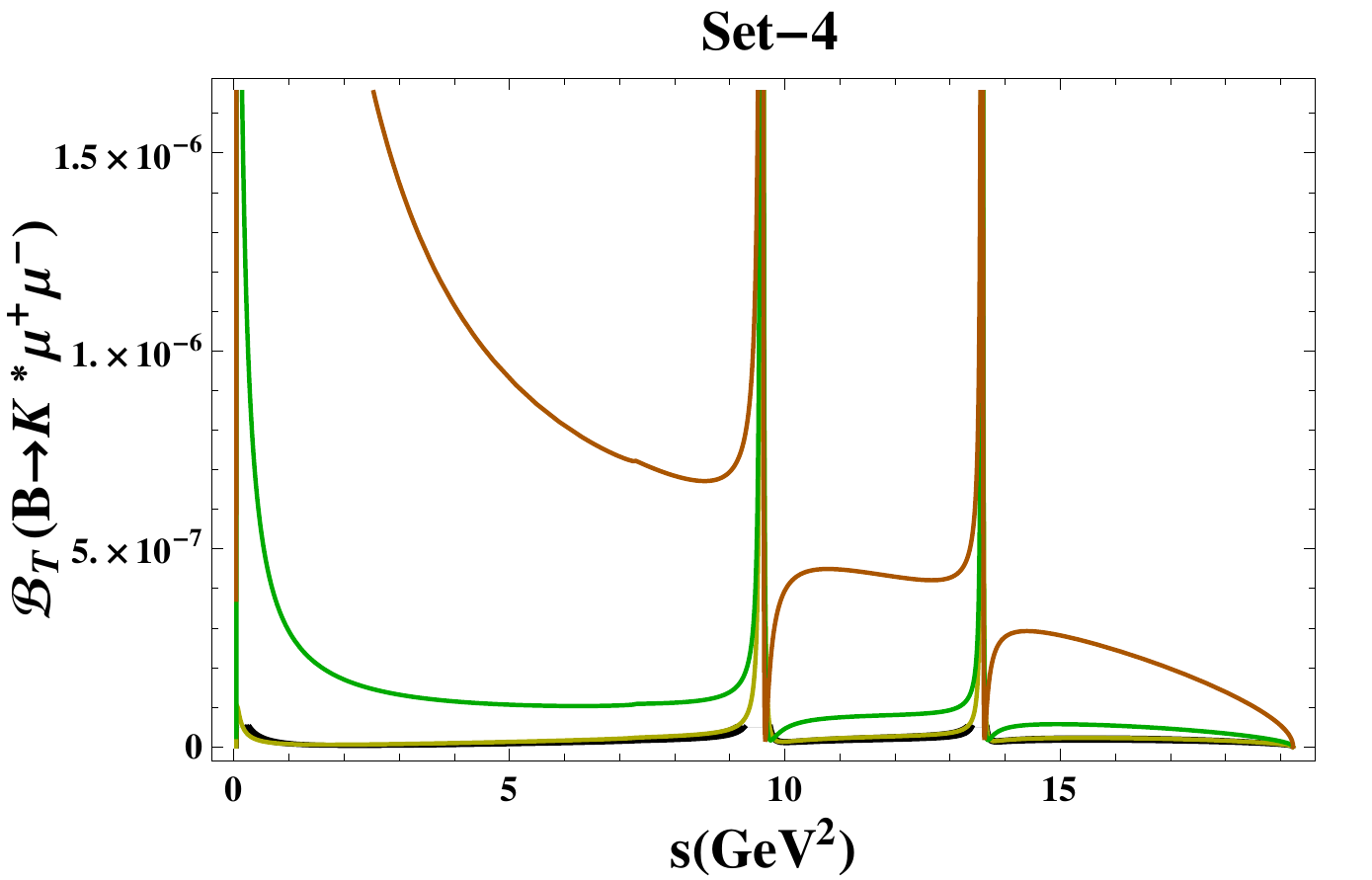}\end{tabular}
\caption{The dependence of transversely polarized branching ratio for $B\to K^{*}\mu^{+}\mu^{-}$ decay on $s$. The line convention is same as in Fig. \ref{longbrmu}.} \label{transbrmu}
\end{figure}

\begin{figure}[ht]
\begin{tabular}{cc}
\centering
\hspace{0.5cm}($\mathbf{a}$)&\hspace{1.2cm}($\mathbf{b}$)\\
\includegraphics[scale=0.55]{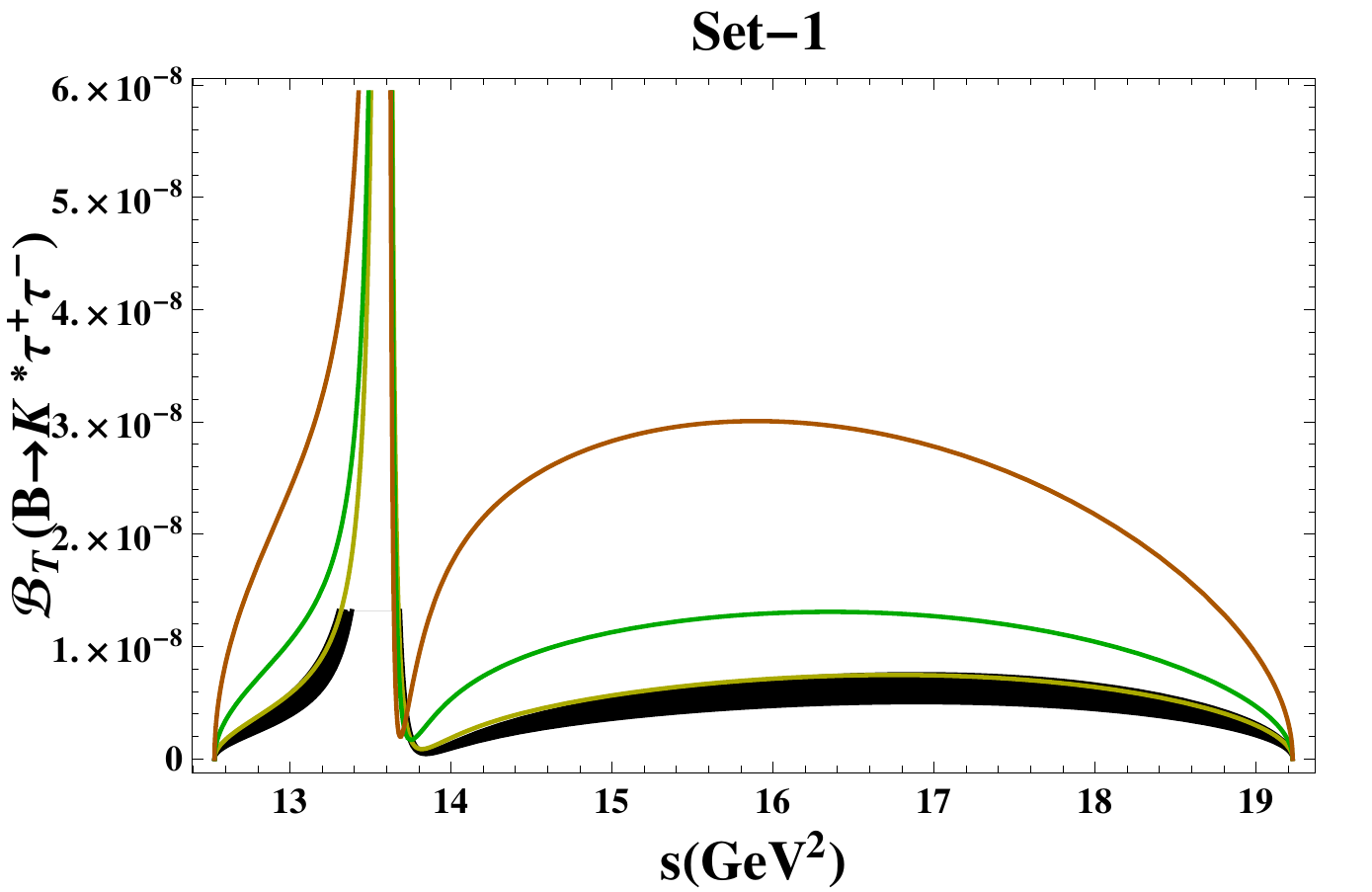}&\includegraphics[scale=0.55]{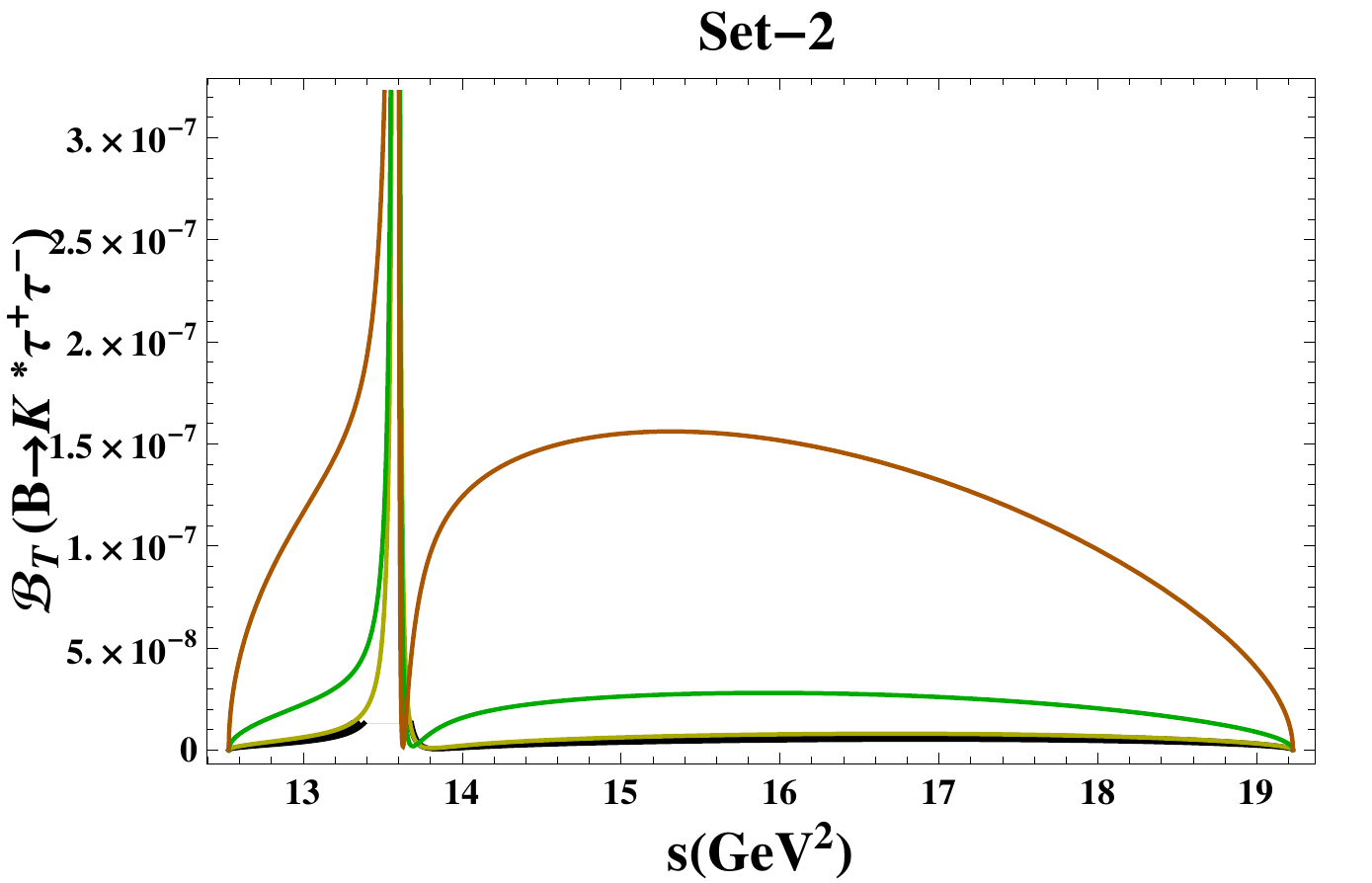}\\
\hspace{0.5cm}($\mathbf{c}$)&\hspace{1.2cm}($\mathbf{d}$)\\
\includegraphics[scale=0.55]{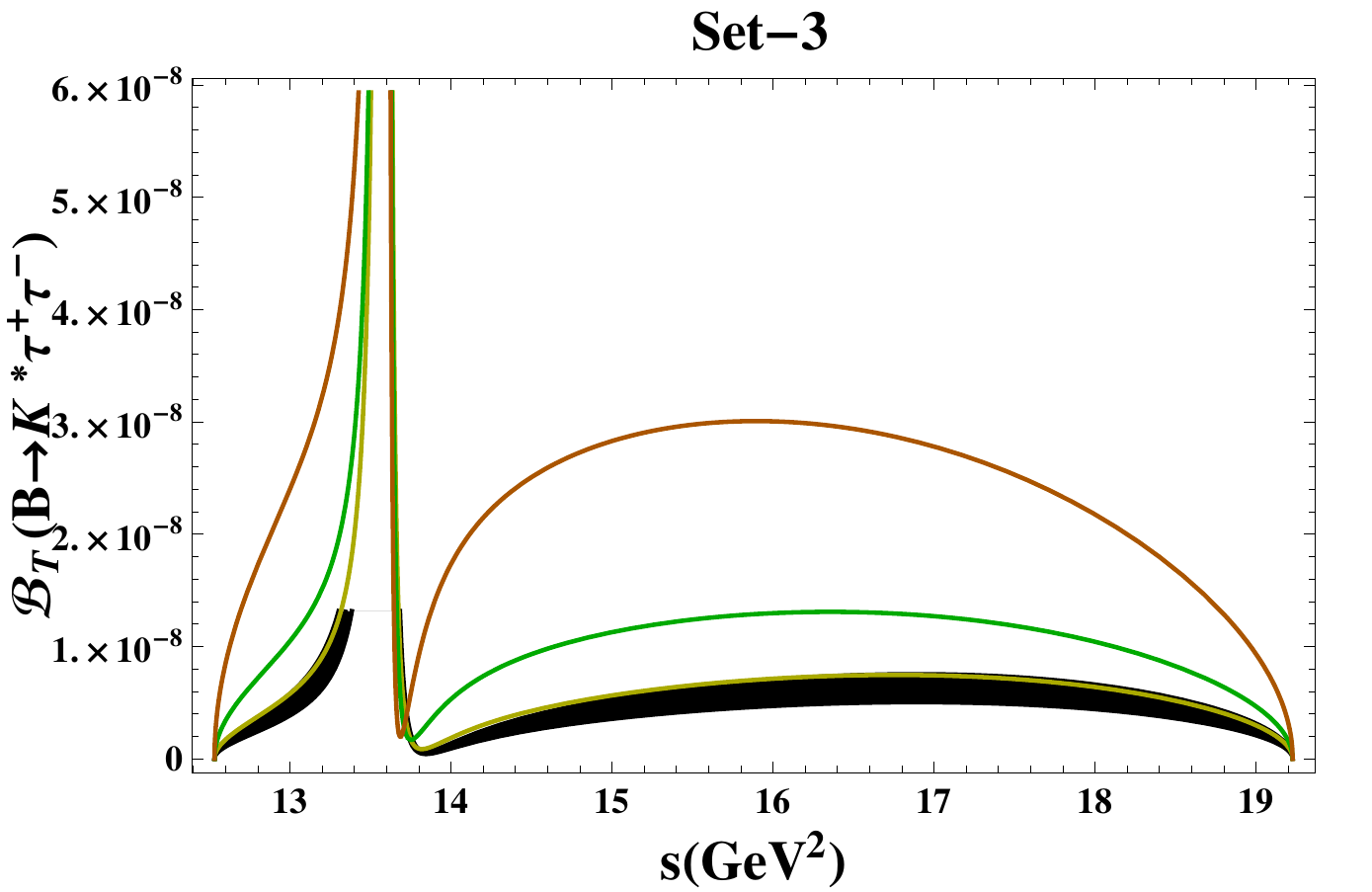}&\includegraphics[scale=0.55]{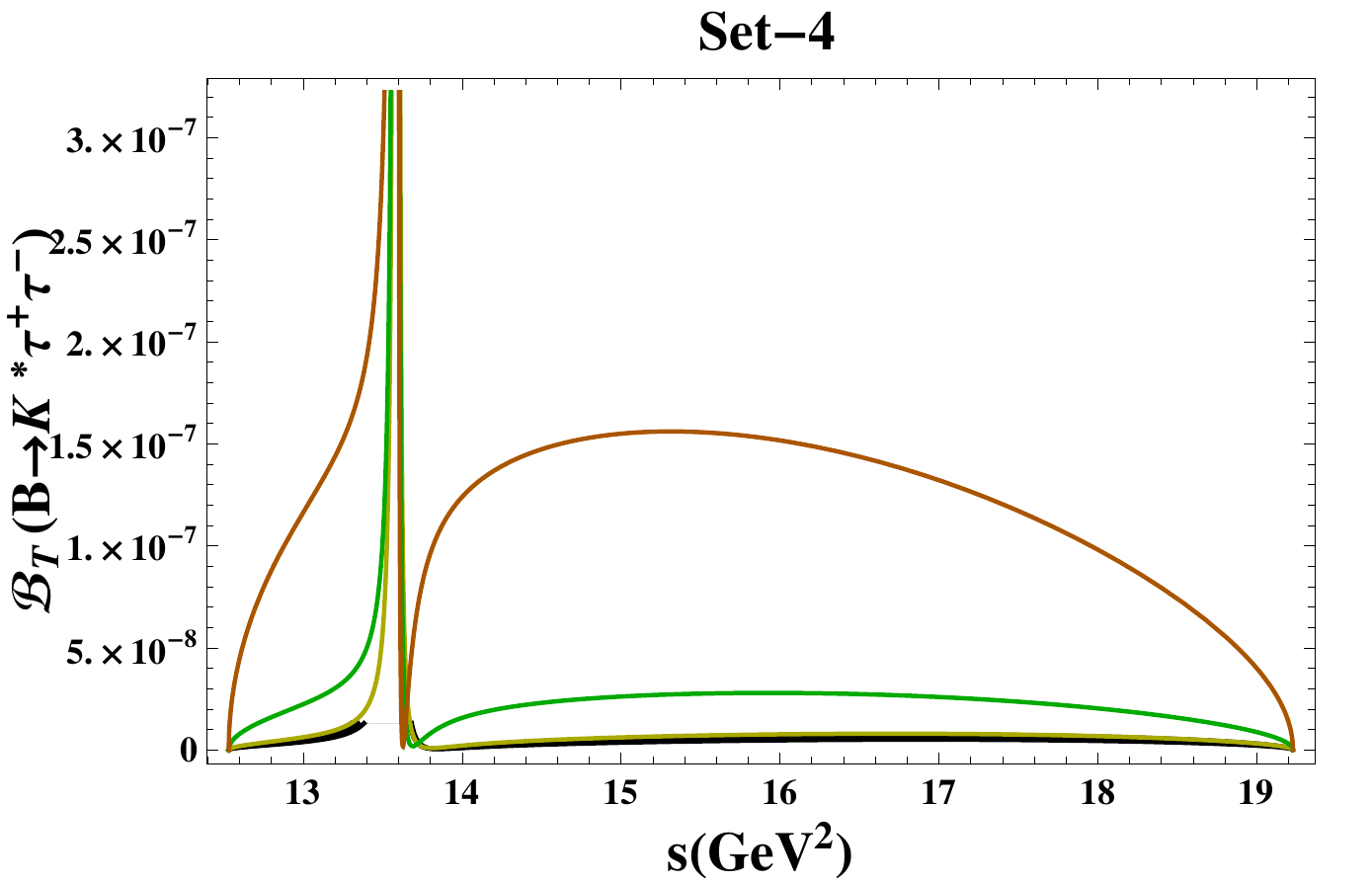}\end{tabular}
\caption{The dependence of transversely polarized branching ratio for $B\to K^{*}\tau^{+}\tau^{-}$ decay on $s$. The line convention is same as in Fig. \ref{longbrmu}.} \label{transbrtau}
\end{figure}

To start with, first we will discuss the polarized branching ratios (${\cal{B}_{L, T}}$) for the decays $B\to K^{\ast}\ell^{+}\ell^{-}$, with $\ell=\mu,\tau$ which we have plotted as a function of $s=q^2$ (GeV$^{2})$ in Figs. \ref{longbrmu}$ - $ \ref{transbrtau}, for different sets of masses given in Table \ref{susy2}. These figures depict the trend of $\mathcal{B}_{L, T}$ both for the cases of muons and tauons as final state leptons. One can notice that the polarized branching ratios are sizeably influenced due to the parametric variation of THDM of type III.  Apart from the graphs we can see that the Wilson coefficients corresponding to the NHBs in the THDM are inversely proportional to the parameter $\lambda_{tt}$ (c.f., Eqs. (\ref{cq1}-\ref{cq5})). Also the Wilson coefficients $C_{Q_1}$ and $C_{Q_2}$ comes with the opposite sign so in this way, their combined contribution is somehow compensated even for the \textit{Case A} where $\lambda_{tt}$ is an order of magnitude smaller than the other two cases. Coming to the Wilson coefficients corresponding to SM, i.e., $C_{7}^{eff}, C_9^{eff}$ and $C_{10}$, the THDM contribution is directly proportional to $\lambda_{tt}$.  Therefore, the effects are expected to be large for large value of $\lambda_{tt}$ which is for the \textit{Case C} and Figs. \ref{longbrmu}$ - $ \ref{transbrtau} justify this fact .

In Fig. \ref{longbrmu} we can see the behaviour of longitudinally polarized branching ratio in $B \to K^* \mu^+ \mu^-$ decay with square of momentum transfer $s$. It can be noticed that for \textit{Case A} the effects of THDM lies inside the uncertainty region of the SM. However, the effects become prominent as we increase the value of $\lambda_{tt}$ and choose different mass sets. In case of mass sets I and III, the most dominant contribution comes for the \textit{Case C} where the longitudinally polarized branching ratio is very distinct from the SM as well as from the \textit{Case A} and \textit{Case B}. The things are even more prominent for mass sets II and IV, where branching ratio for \textit{Case B} and \textit{Case C} are an order of magnitude larger than the corresponding SM and \textit{Case A} values. This can also be seen in Table \ref{longitudinal-table}, where we have given the integrated values of above mentioned observables in the limit that lies below the resonance region. Similar effects can also be seen in $B \to K^* \tau^+ \tau^-$ decay (c.f. , Fig. \ref{longbrtau}).

Likewise, we have plotted the transverse polarized branching ratio of $B \to K^* \ell^+ \ell^-$ against $s$ in Figs. \ref{transbrmu} and \ref{transbrtau} for the cases when we $\mu$'s and $\tau$'s as final state leptons. In Eq. (\ref{ATfunction}) one can notice that the amplitude corresponding to transversely polarized branching ratio $(\mathcal{A}_T)$ does not depend on auxilary functions $f_7$ and $f_8$ and so on the neutral Higgs boson contributions. Therefore, the entire deviations from the SM results come through the Wilson coefficients $C_{7}^{eff}, C_9^{eff}$ and $C_{10}$ that are directly proportional to $\lambda_{tt}$. Hence, the most prominent results are expected for the  \textit{Case B}  and  \textit{Case C} and Figs. \ref{transbrmu} and \ref{transbrtau} justify this claim. It can also be seen in Table \ref{transverse-table} where the values of transverse polarized branching ratio  differs by the orders of magnitude for mass Sets II and IV. Therefore, the measurement of these observables will help us to restrict the parametric space of type III of the THDM.

\begin{table}[ht]
\centering
\begin{tabular}{lcccc}
\hline\hline
$\mathcal{B}_L$ &  & Case A& Case B & Case C \\
\hline
SM &  $4.56^{+1.02}_{-1.16}\times10^{-7}$& $$&  & \\
Set I & &$4.90\times10^{-7}$ &$6.59\times10^{-7}$  & $1.15\times10^{-6}$\\
Set II & &  $5.05\times10^{-7}$&$1.09\times10^{-6}$& $4.30\times10^{-6}$\\
Set III & &$4.90\times10^{-7}$ &$6.59\times10^{-7}$  & $1.15\times10^{-6}$\\
Set IV & & $5.05\times10^{-7}$&$1.09\times10^{-6}$& $4.30\times10^{-6}$\\
\hline\hline
\end{tabular}
\caption{The longitudinally polarized branching ratio for $B\to K^{\ast}\mu^{+}\mu^{-}$ decay in the SM and THDM with different set of masses. The limit of integration on $q^2$ is set to be below the resonance region.}\label{longitudinal-table}
\end{table}

\begin{table}[ht]
\centering
\begin{tabular}{lcccc}
\hline\hline
$\mathcal{B}_T$ &  & Case A& Case B & Case C \\
\hline
SM &  $1.52^{0.31}_{-0.29}\times10^{-7}$& $$&  & \\
Set I & &$1.67\times10^{-7}$ &$4.57\times10^{-7}$  & $2.10\times10^{-6}$\\
Set II & & $1.78\times10^{-7}$&$1.87\times10^{-6}$& $2.00\times10^{-5}$\\
Set III & &$1.67\times10^{-7}$ &$4.57\times10^{-7}$  & $2.10\times10^{-6}$\\
Set IV & & $1.78\times10^{-7}$&$1.87\times10^{-6}$& $2.00\times10^{-5}$\\
\hline\hline
\end{tabular}
\caption{The transversely polarized branching ratio for $B\to K^{\ast}\mu^{+}\mu^{-}$ decay in the SM and THDM with different set of masses. The limit of integration on $q^2$ is set to be below the resonance region.}\label{transverse-table}
\end{table}

Another interesting observables in exclusion $B$ meson decays are the various lepton polarization asymmetries. Contrary to the branching ratio, these are less influenced by the uncertainties coming through different inputs where the most important are the form factors. Keeping this in view, below we are going to discuss the longitudinal and normal lepton polarization asymmetries in $B\to K^{\ast}\ell^{+}\ell^{-}$ decay.

\begin{figure}[ht]
\begin{tabular}{cc}
\centering
\hspace{0.5cm}($\mathbf{a}$)&\hspace{1.2cm}($\mathbf{b}$)\\
\includegraphics[scale=0.55]{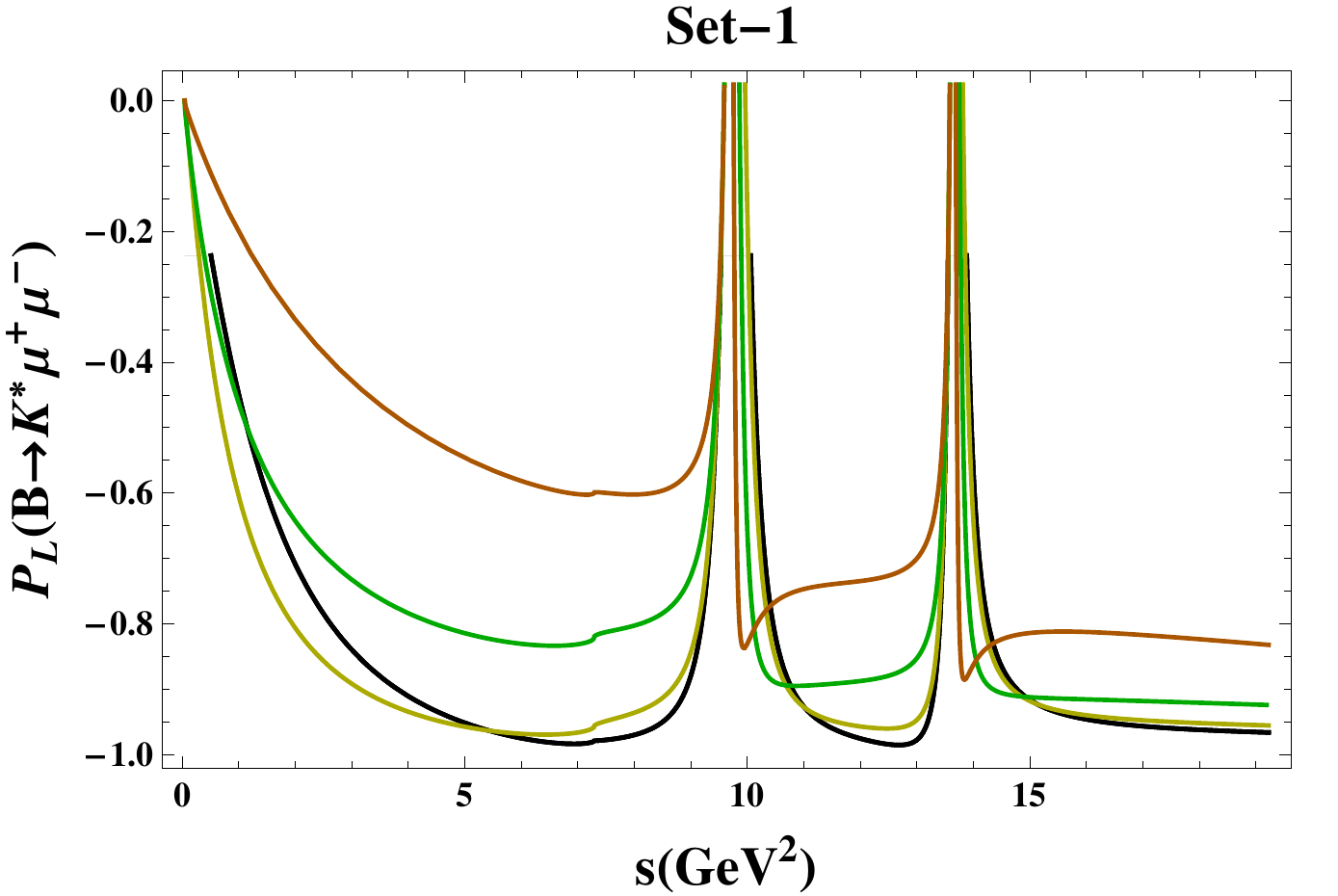}&\includegraphics[scale=0.55]{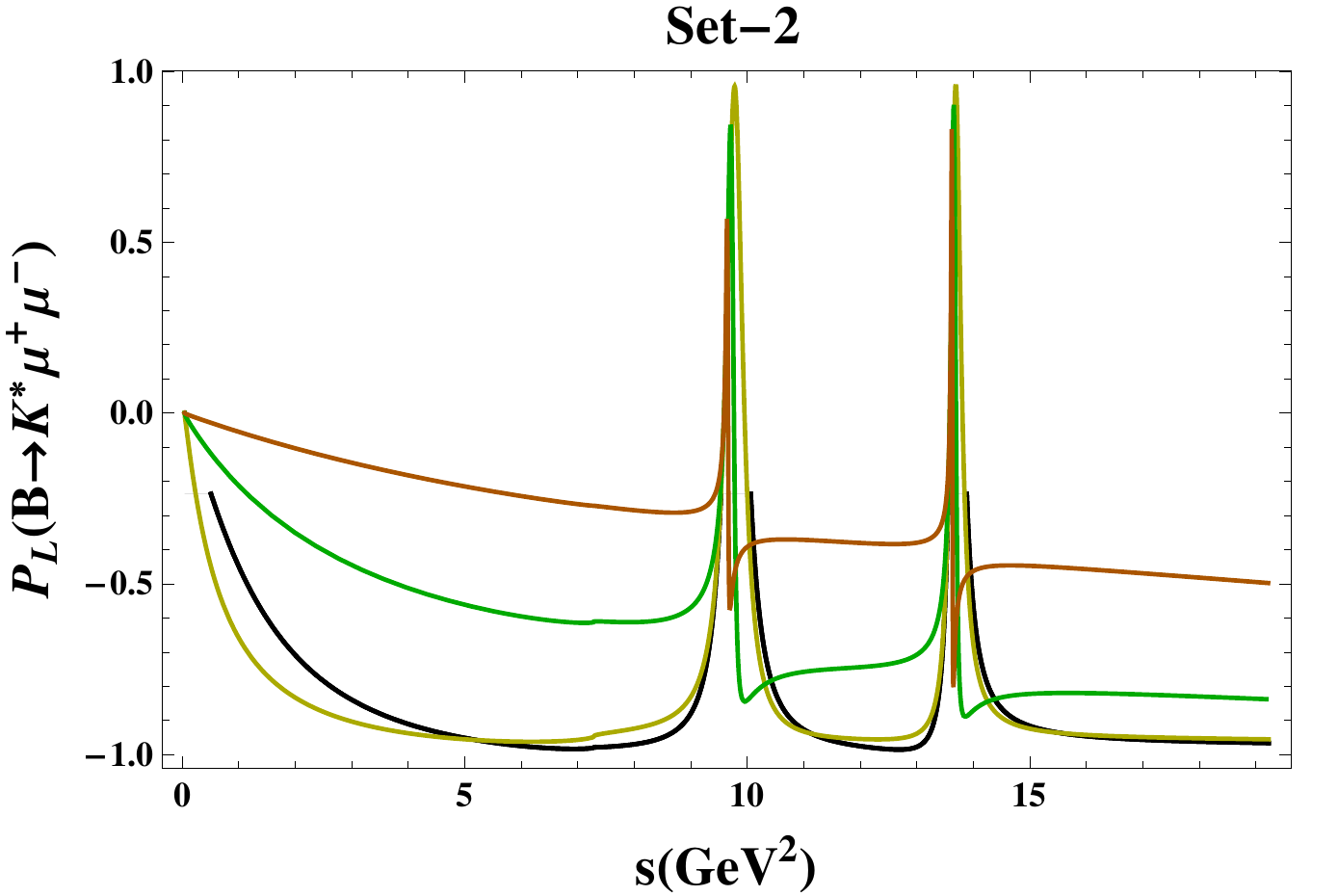}\\
\hspace{0.5cm}($\mathbf{c}$)&\hspace{1.2cm}($\mathbf{d}$)\\
\includegraphics[scale=0.55]{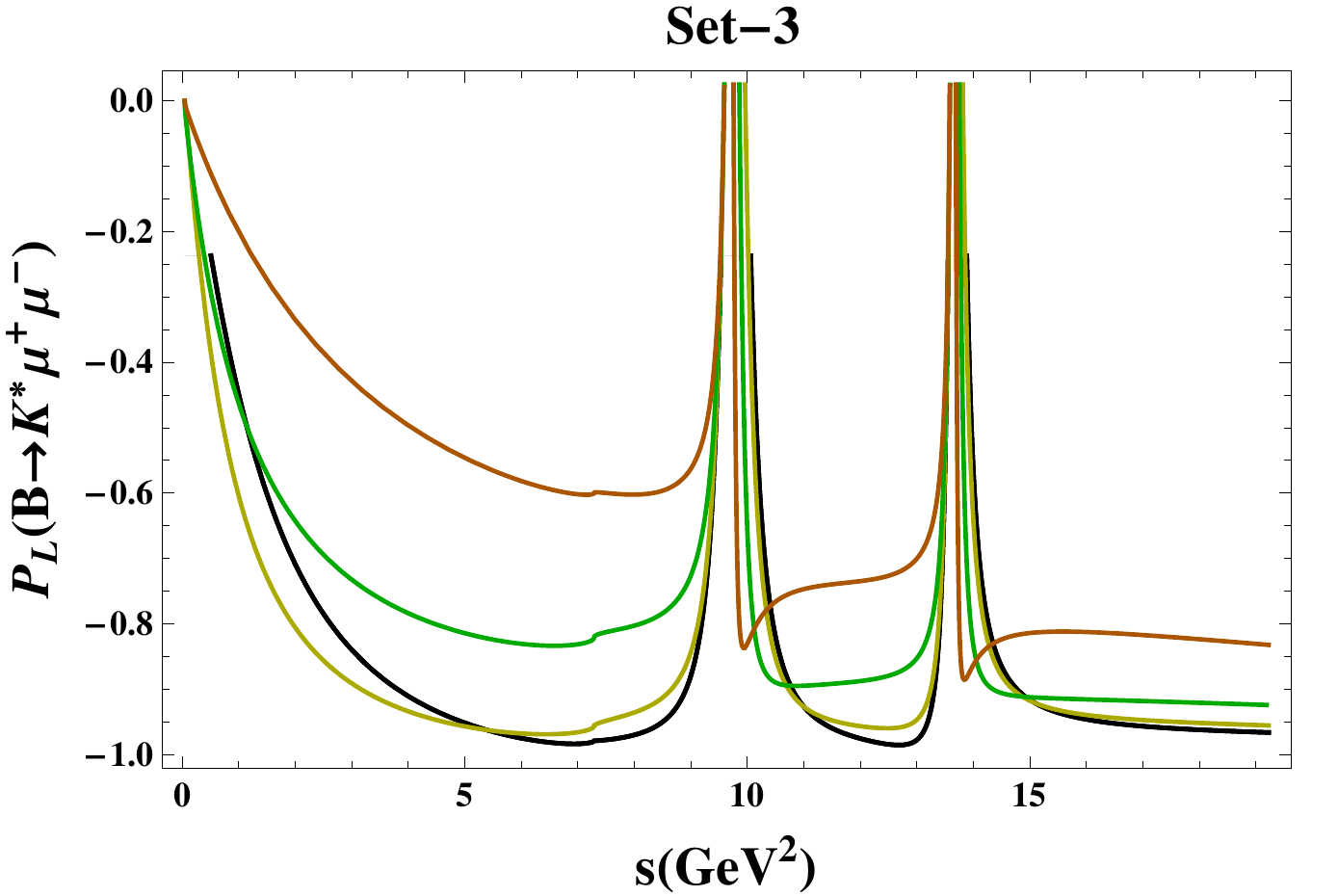}&\includegraphics[scale=0.55]{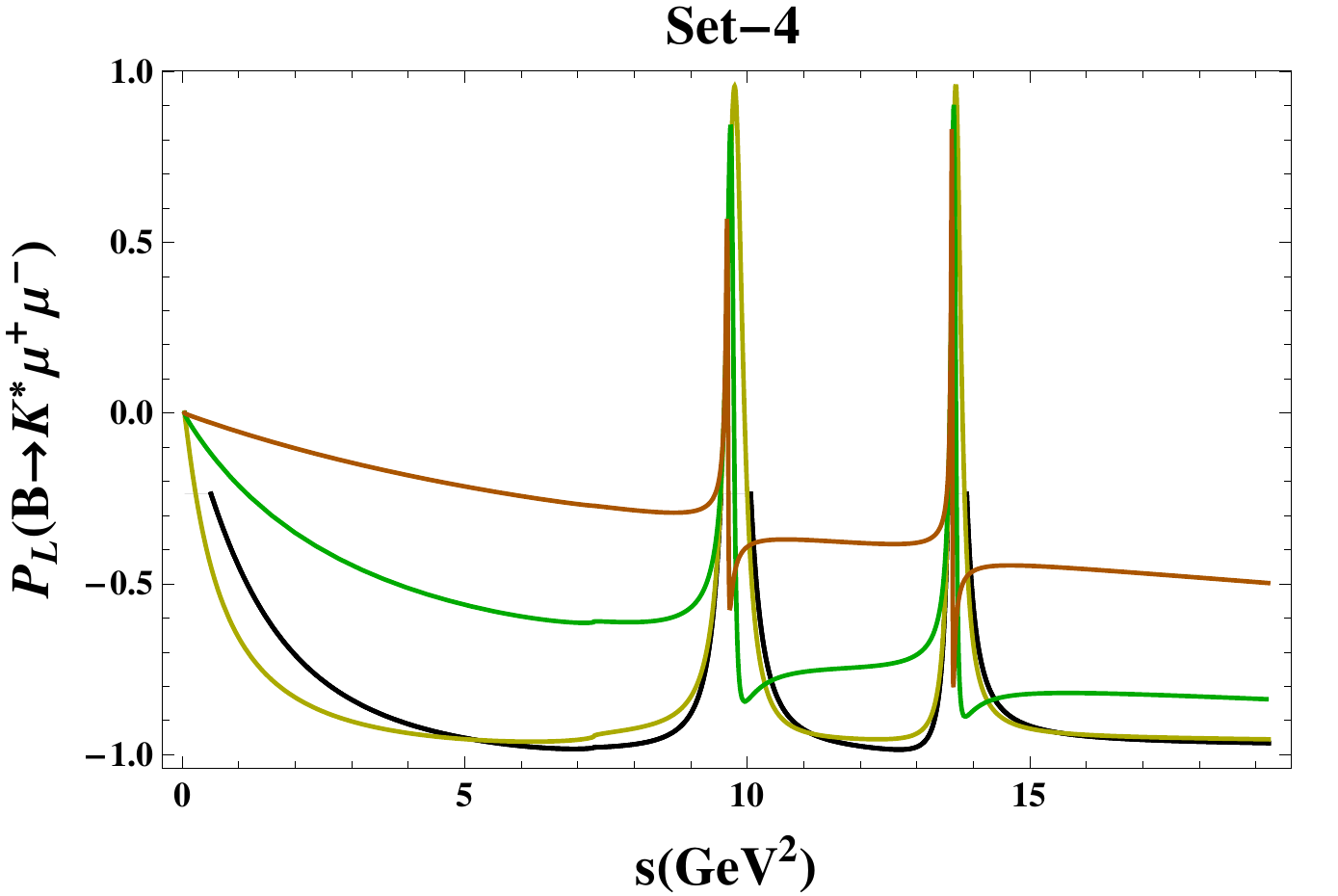}\end{tabular}
\caption{The dependence of longitudinal lepton polarization  for $B\to K^{*}\mu^{+}\mu^{-}$ decay on $s$. The legends are same as in Fig. \ref{longbrmu}.} \label{brm2}
\end{figure}

First, we will discuss the longitudinal lepton polarization asymmetry ($P_L$) whose expression is given in Eq. (\ref{long-polarization}). It can be noticed that, this asymmetry depends on different combinations of the Wilson Coefficients that are encoded in the auxiliary functions $f_1, . . ., f_8$. Therefore, one can expect large dependency of these asymmetries on the various parameters of the THDM, hence, making this observable fertile to extract the NP. In Eq. (\ref{long-polarization}), the last term which involve the contributions from the NHBs is the mass of lepton suppressed compared to other terms, therefore, it will be less important for $\mu$'s as compared to the $\tau$'s as final state leptons. 
The dependence of $P_L$ in $B \to K^* \mu^+\mu^-$ decay on $s$ is depicted in Fig. \ref{brm2} for different values of $\lambda_{tt}$ along with different mass sets. It can be noticed that the value of $P_L$ differs significantly from the SM results for almost all the cases. The derivations are most prominent in \textit{Case C} where $\lambda_{tt}$ is an order of magnitude larger than the \textit{Case A}. In addition to the parameters $\lambda_{tt}$ and $\lambda_{bb}$, the Wilson coefficients $C_{7}^{eff}, C_9^{eff}$ and $C_{10}$ also depend on a parameter $y=\frac{m^2_{t}}{m_{H^\pm}^2}$, hence different choice of the mass of charged Higgs boson will give us some difference in the results. This can be easily noticed in Fig.\ref{brm2} for different mass sets and also in Table \ref{PL-table}. Here, we can see that for $\lambda_{tt}=0.3$, the average value of $P_L$ shifted from $-0.771$ to $-0.339$ and $-0.10$ for the mass Sets I, III and Sets II, IV, respectively.

\begin{figure}[ht]
\begin{tabular}{cc}
\centering
\hspace{0.5cm}($\mathbf{a}$)&\hspace{1.2cm}($\mathbf{b}$)\\
\includegraphics[scale=0.55]{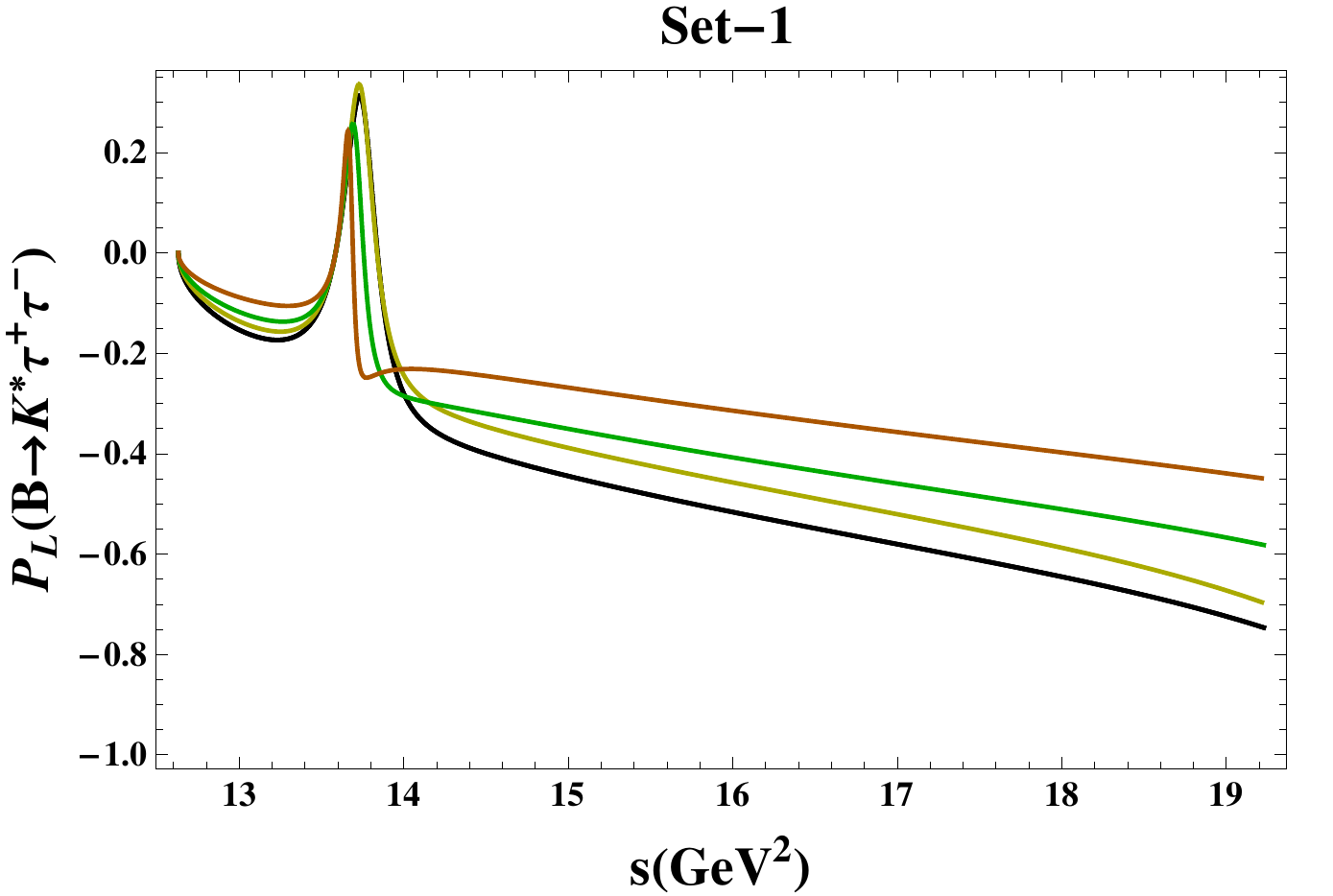}&\includegraphics[scale=0.55]{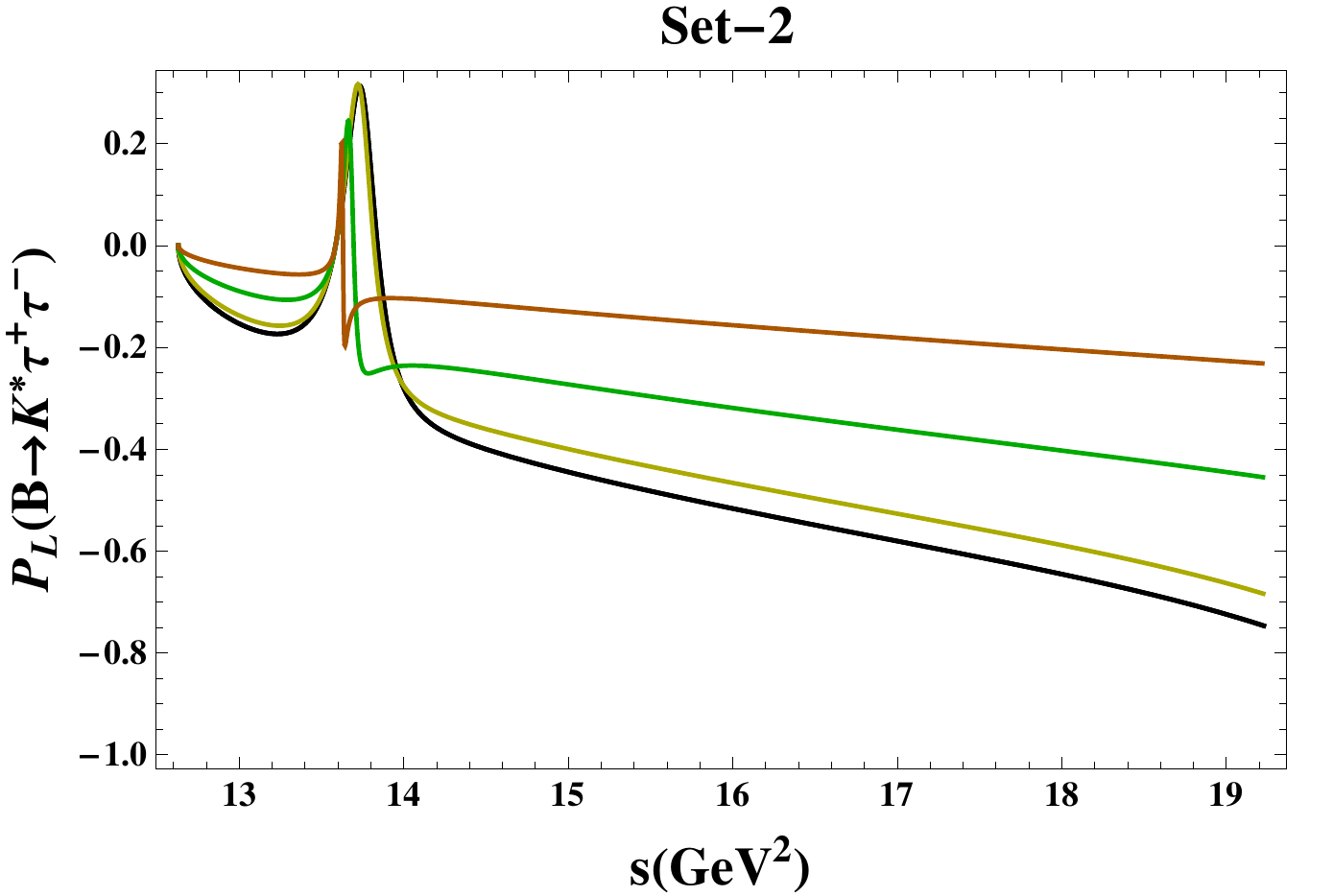}\\
\hspace{0.5cm}($\mathbf{c}$)&\hspace{1.2cm}($\mathbf{d}$)\\
\includegraphics[scale=0.55]{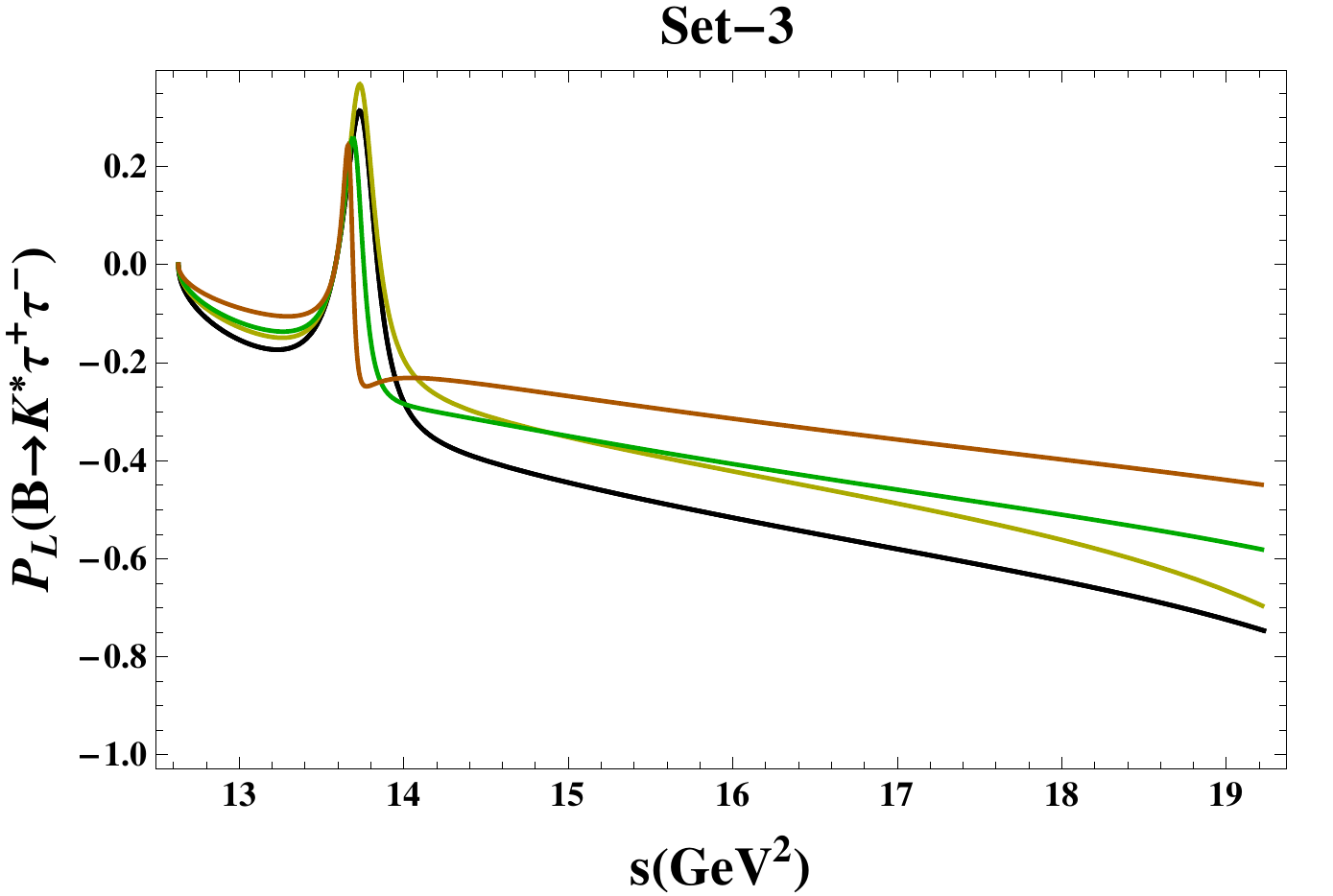}&\includegraphics[scale=0.55]{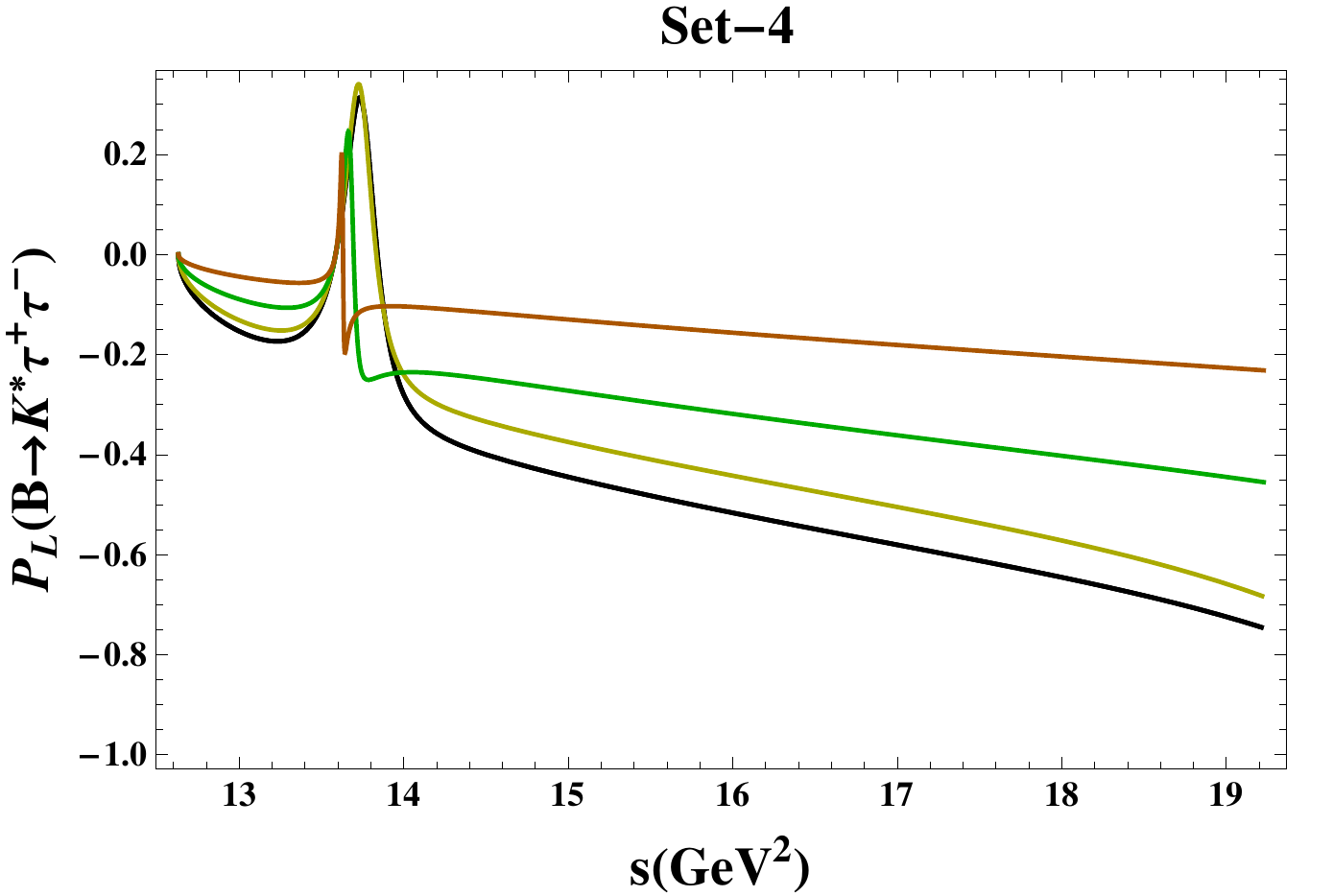}\end{tabular}
\caption{The dependence of longitudinal lepton polarization  for $B\to K^{*}\tau^{+}\tau^{-}$ decay on $s$. The legends are same as in Fig. \ref{longbrmu}.} \label{brm3}
\end{figure}

In Fig. \ref{brm3} we have plotted the trend of $P_L$ for $B \to K^* \tau^+ \tau^-$ decay with the square of momentum transfer $s$ for different choices of the parameters of the THDM of type III. Contrary to the $\mu$'s as final state leptons, in the present case, the effects due to different choices of mass sets are more pertinent. The one obvious reason is that the suppression due to mass of final state lepton in the last term of Eq. (\ref{long-polarization}) is somehow lifted in comparison to the previous case. We can see that the maximum value of $P_L$ shift in mass Sets I and III are of the factor of 3, where as for the Sets II and IV this shift is a factor of 4. Therefore, the measurement of this observable may help us to distinguish between different ranges of mass parameters in THDM of type III.

\begin{table}[ht]
\centering
\begin{tabular}{lcccc}
\hline\hline
$\langle  P_L\rangle$ &  & Case A& Case B & Case C \\
\hline
SM &  $-0.771$& $$&  & \\
Set I & &$-0.808$ &$-0.644$  & $-0.339$\\
Set II & &  $-0.817$&$-0.335$& $-0.10$\\
Set III & &$-0.808$ &$-0.644$  & $-0.339$\\
Set IV & & $-0.817$&$-0.335$& $-0.10$\\
\hline\hline
\end{tabular}
\caption{The average value of longitudinally polarized lepton asymmetry for $B\to K^{\ast}\mu^{+}\mu^{-}$ decay in the SM and THDM with different set of masses. The limit of integration on $q^2$ is set to be below the resonance region.}\label{PL-table}
\end{table}

\begin{figure}[ht]
\begin{tabular}{cc}
\centering
\hspace{0.5cm}($\mathbf{a}$)&\hspace{1.2cm}($\mathbf{b}$)\\
\includegraphics[scale=0.55]{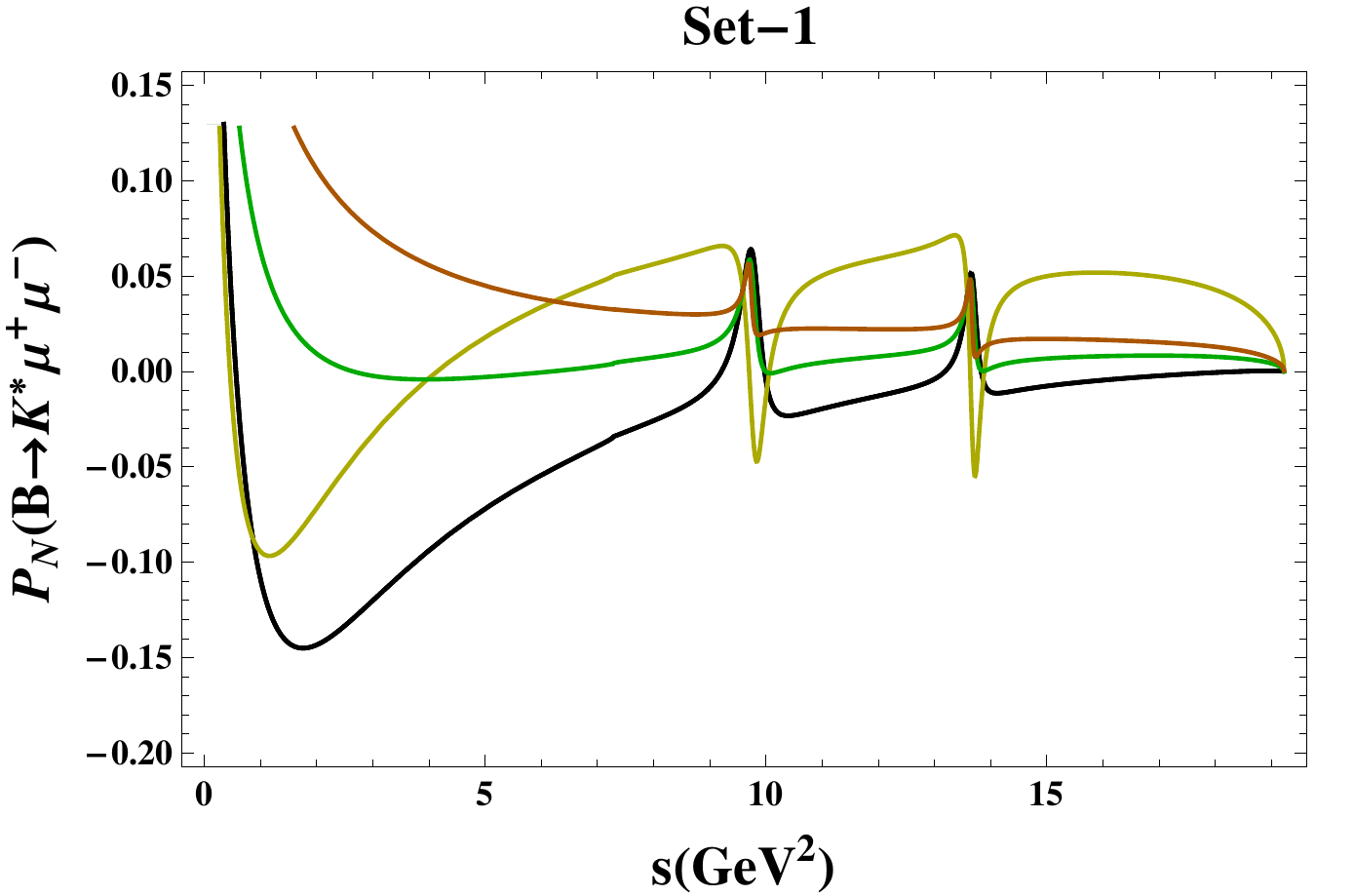}&\includegraphics[scale=0.55]{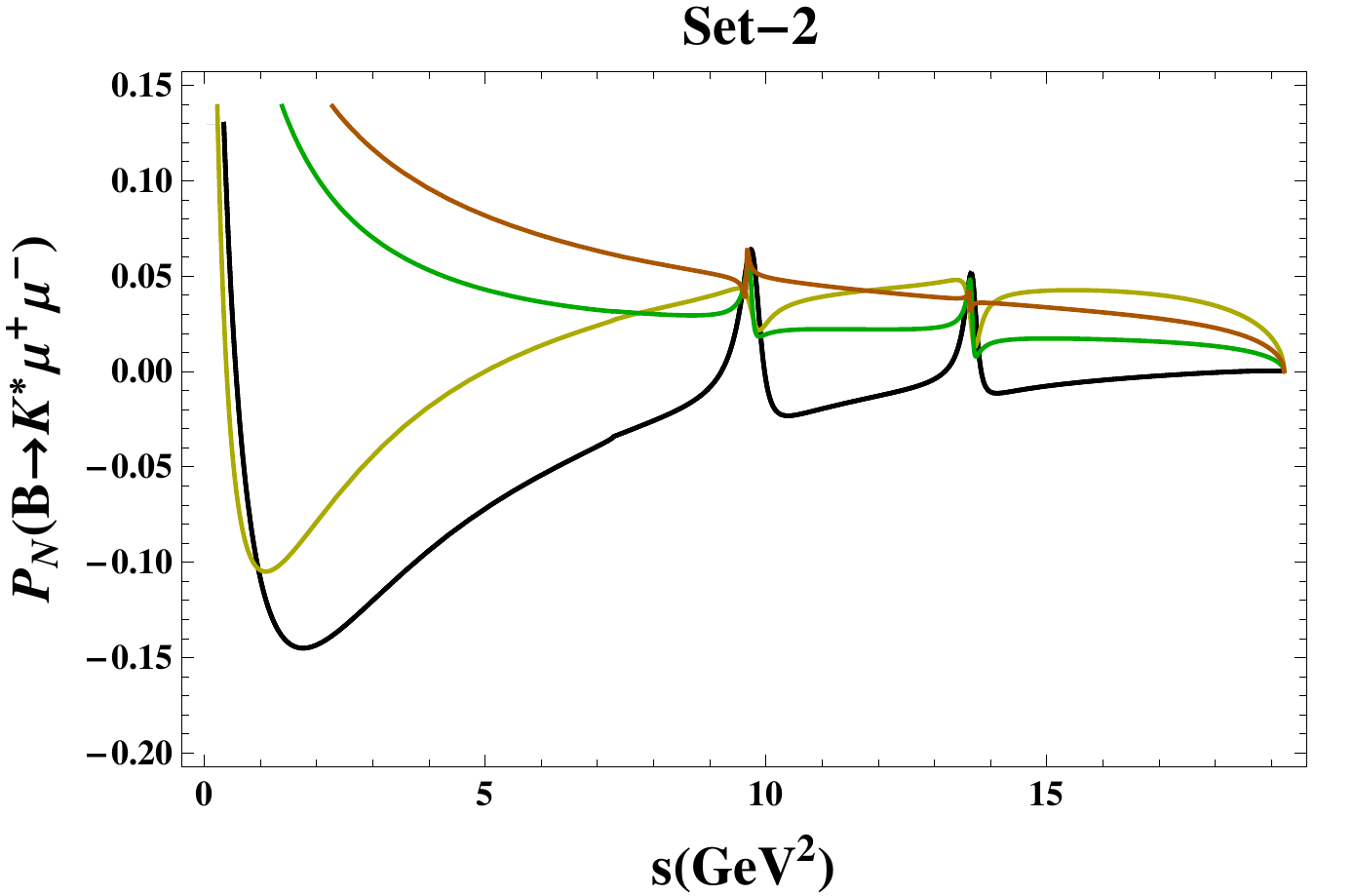}\\
\hspace{0.5cm}($\mathbf{c}$)&\hspace{1.2cm}($\mathbf{d}$)\\
\includegraphics[scale=0.55]{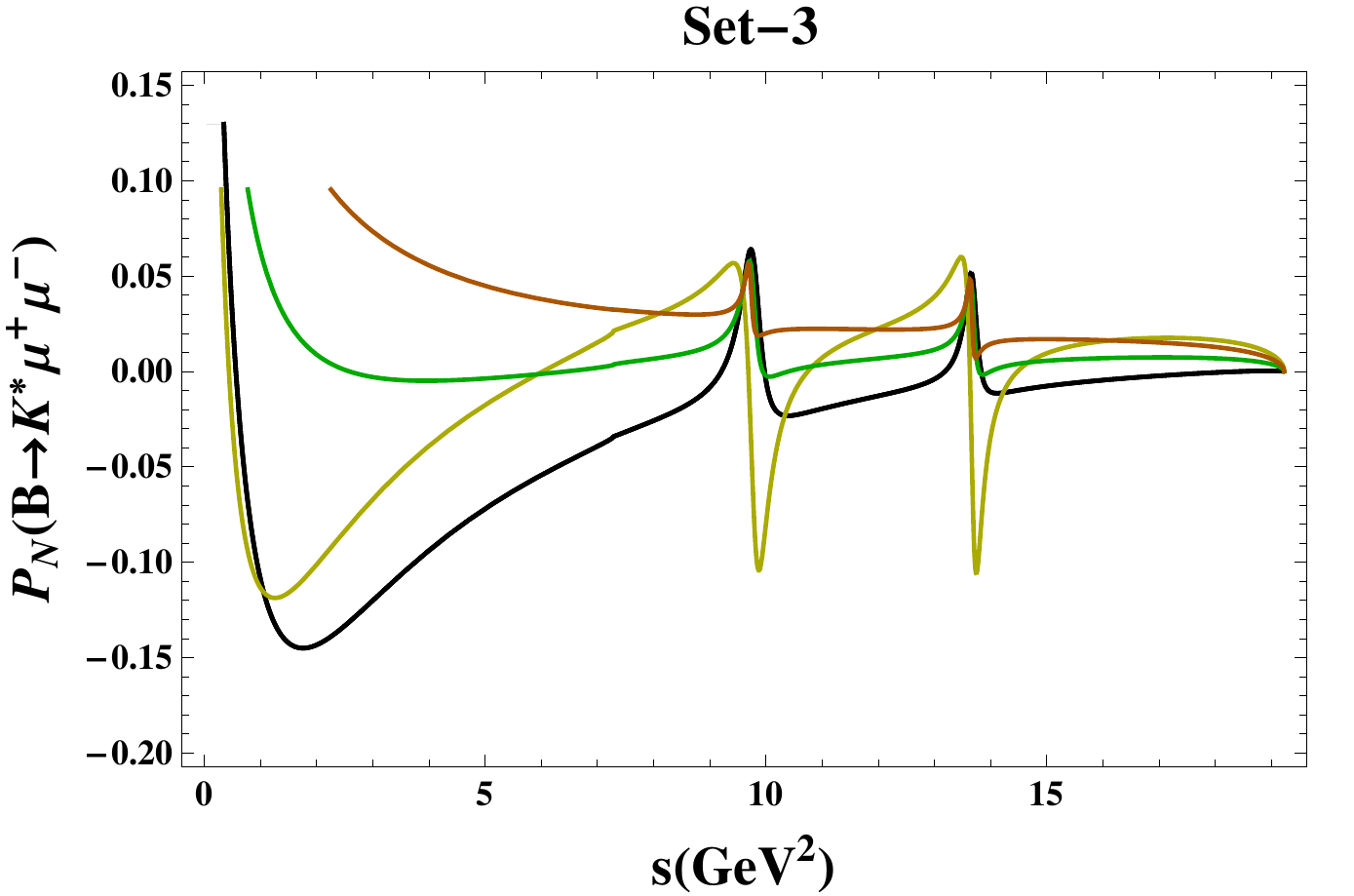}&\includegraphics[scale=0.55]{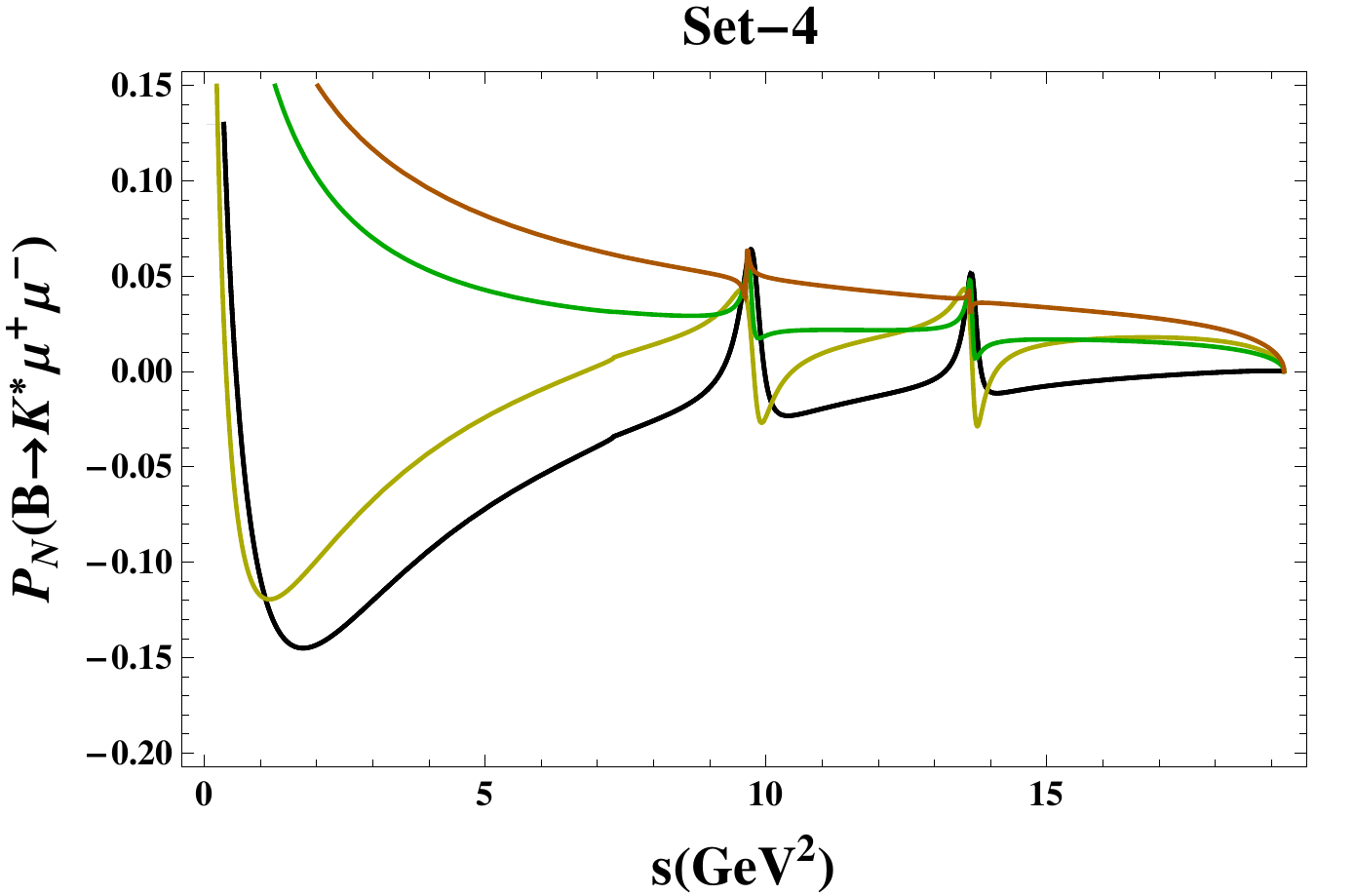}\end{tabular}
\caption{The dependence of normal lepton polarization  for $B\to K^{*}\mu^{+}\mu^{-}$ decay on $s$. The legends are same as in Fig. \ref{longbrmu}.} \label{brm4}
\end{figure}

\begin{figure}[ht]
\begin{tabular}{cc}
\centering
\hspace{0.5cm}($\mathbf{a}$)&\hspace{1.2cm}($\mathbf{b}$)\\
\includegraphics[scale=0.55]{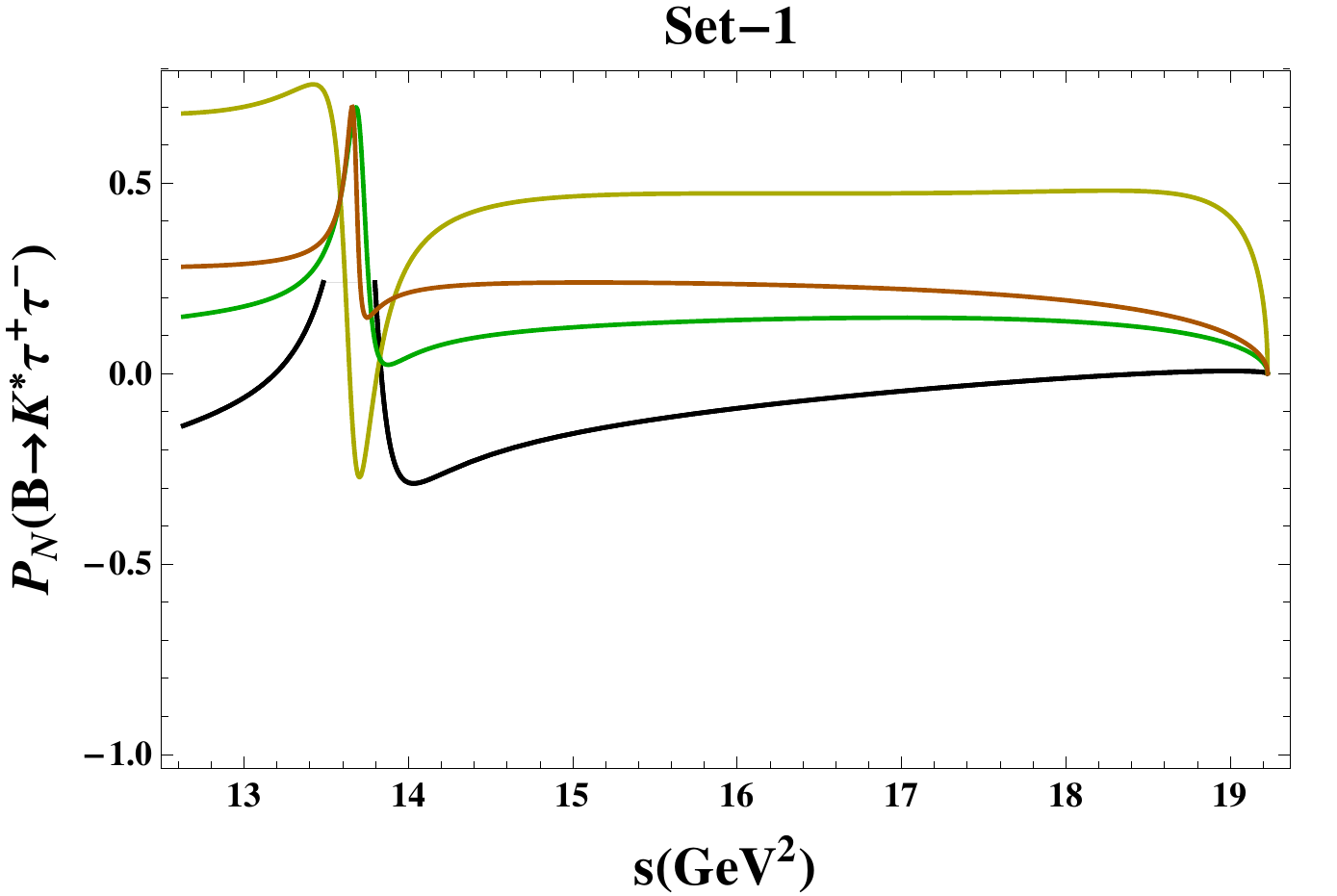}&\includegraphics[scale=0.55]{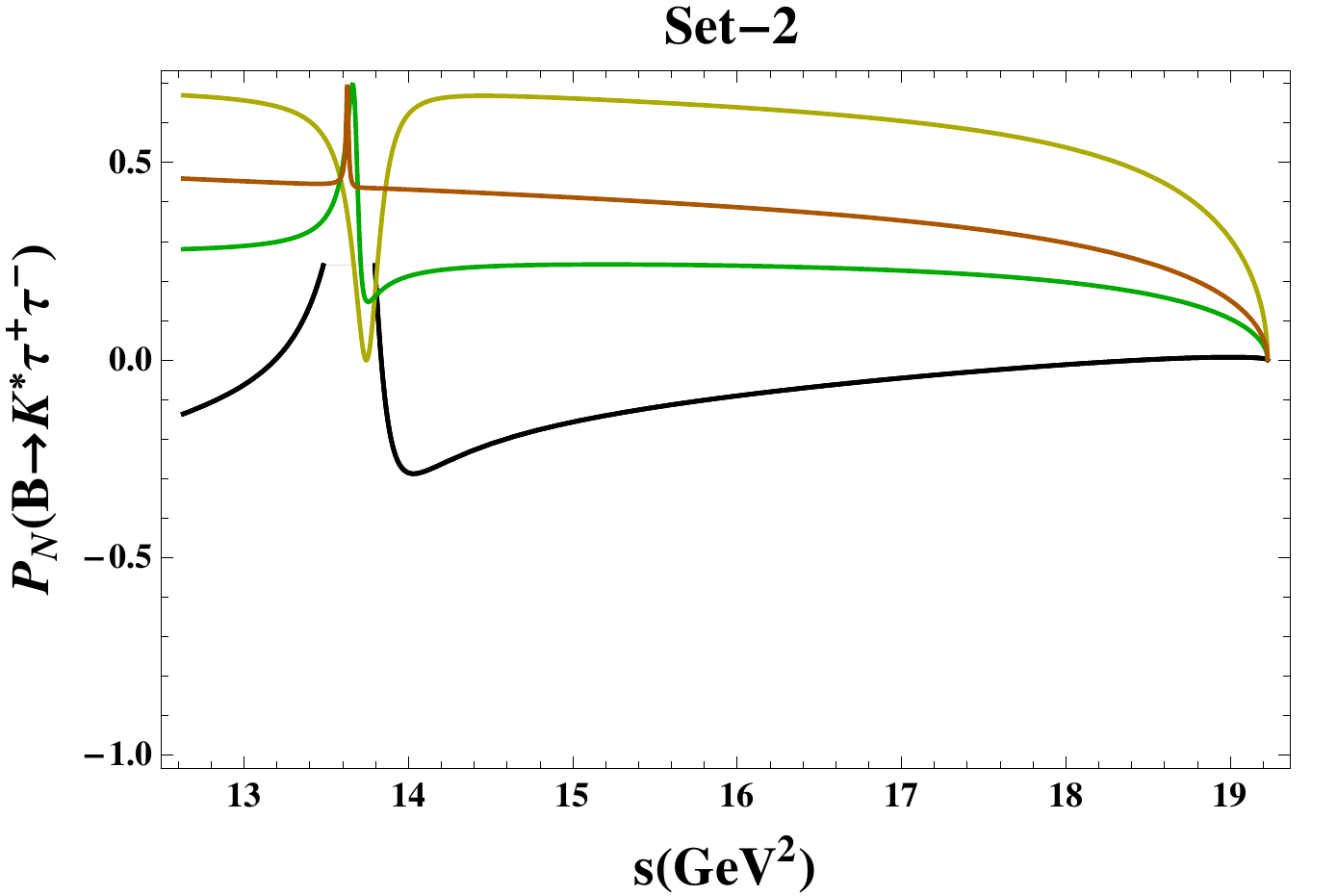}\\
\hspace{0.5cm}($\mathbf{c}$)&\hspace{1.2cm}($\mathbf{d}$)\\
\includegraphics[scale=0.55]{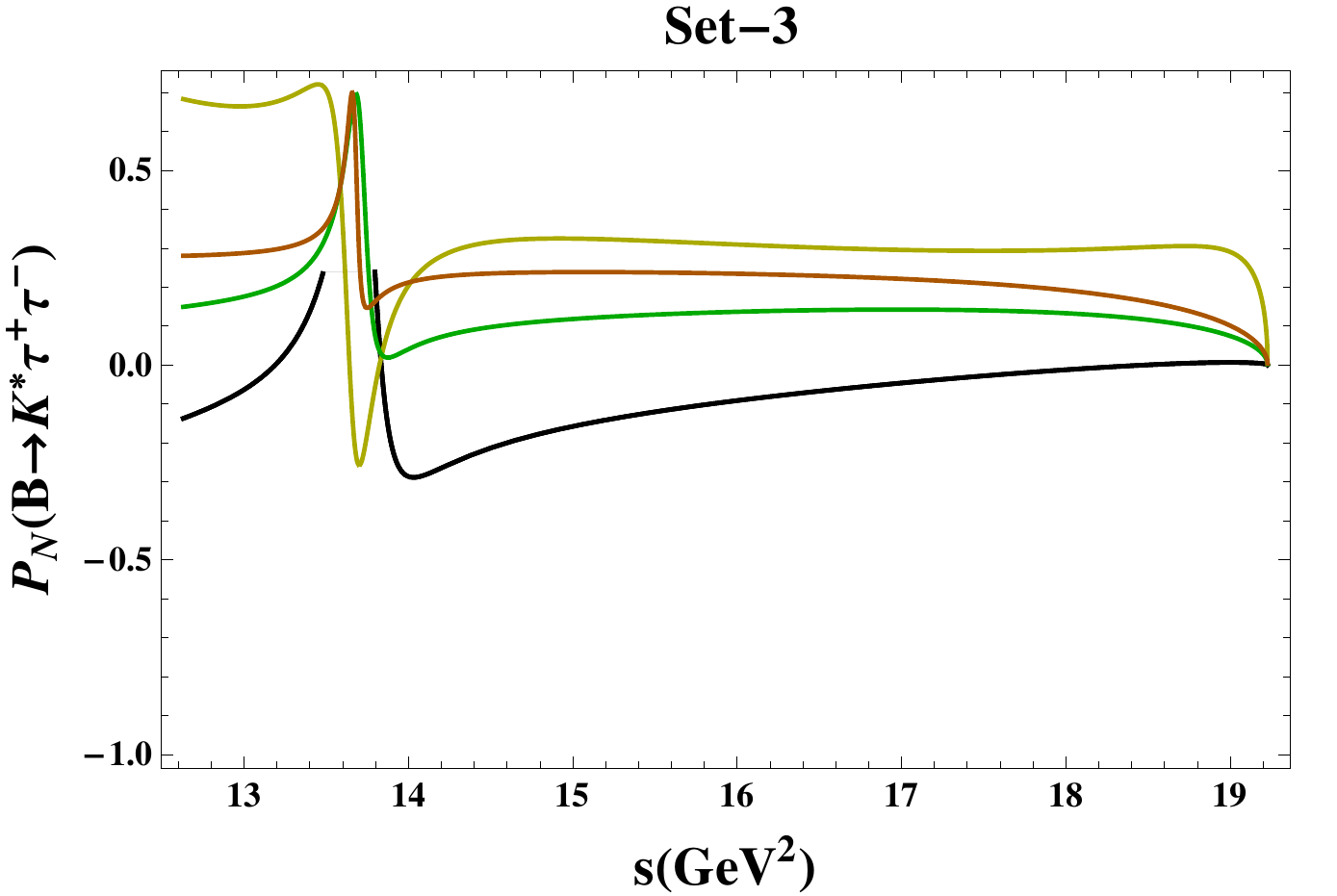}&\includegraphics[scale=0.55]{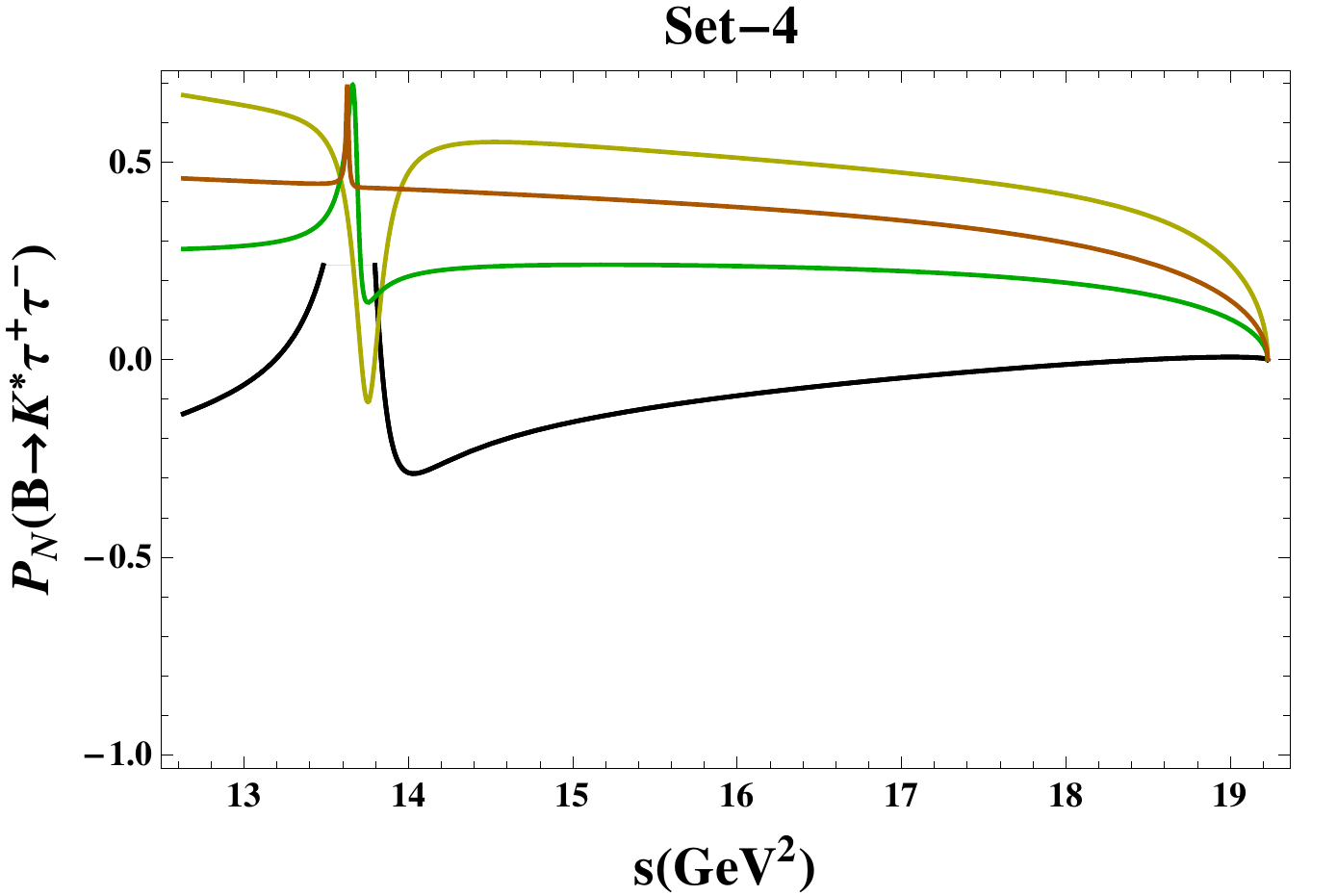}\end{tabular}
\caption{The dependence of normal lepton polarization  for $B\to K^{*}\tau^{+}\tau^{-}$ decay on $s$. The legends are same as in Fig. \ref{longbrmu}.} \label{brm5}
\end{figure}

Figs. \ref{brm4} and \ref{brm5}  show the dependence of normal lepton polarization asymmetries with the square of momentum transfer for the above $B \to K^* \mu^+ \mu^-$ and $B \to K^* \tau^+ \tau^-$ decays, respectively.  Contrary to the case of longitudinal lepton polarization (c.f. Eq. (\ref{long-polarization})), where most of the terms are positive and the one which is negative is mass of lepton suppressed, in the present case, it is evident from Eq. (\ref{norm-polarization}) that we have both positive and negative terms in the expression. It is, therefore, expected that at certain value of $q^2$ this asymmetry will cross zero in the SM (even below the resonance region) and Fig. \ref{brm4} justify this fact. This zero crossing can also be seen for the \textit{Case A} of the THDM, where the value of $P_N$ remains positive for the \textit{Case B} and \textit{Case C}. Compared to the $\langle P_L\rangle$ the value of $\langle P_N\rangle$ has suppressed because of its proportionality to mass of lepton (c.f. Eq. (\ref{norm-polarization})). 

In Table \ref{PN-table}, we have given the values of $\langle P_N \rangle$ after making integration on the square of momentum transfer $s$ in the range $s_{min} \leq s \leq m^2_{\psi}$, i.e., below the resonance region for $B \to K^* \mu^+ \mu^-$. It can be noticed that the $\langle P_N \rangle$ is negative in the SM, where as it is possible for all the cases of THDM of type III. For the \textit{Case} C the value of $\langle P_N \rangle$is positive and an order of magnitude different from its SM value. Therefore, it will be interesting to look for this observable at the ongoing and future experiments.

\begin{table}[ht]
\centering
\begin{tabular}{lcccc}
\hline\hline
$\langle  P_N\rangle$ &  & Case A& Case B & Case C \\
\hline
SM &  $-0.013$& $$&  & \\
Set I & &$0.04$ &$0.06$  & $0.19$\\
Set II & &  $0.017$&$0.18$& $0.28$\\
Set III & &$0.04$ &$0.06$  & $0.19$\\
Set IV & &$0.017$&$0.18$& $0.28$\\
\hline\hline
\end{tabular}
\caption{The average value of longitudinally polarized lepton asymmetry for $B\to K^{\ast}\mu^{+}\mu^{-}$ decay in the SM and THDM with different set of masses. The limit of integration on $q^2$ is set to be below the resonance region.}\label{PN-table}
\end{table}

We now discuss the dependence of transverse polarization asymmetries on square of momentum transfer for the decays $B\to K^{\ast}\ell^{+}\ell^{-}$. It can be seen from Eq. (\ref{Transverse-polarization}), that it is proportional to the imaginary parts of the Wilson coefficients which are too small both in the SM as well in the THDM. Therefore, the value of $P_T$ is too small to be measured experimentally.

\section{Conclusion}\label{con}

This study on the rare $B\to K^{\ast}\ell^{+}\ell^{-}$ decays, with $\ell=\mu$, $\tau$, focus on the analysis of polarized branching ratio ($\mathcal{B}_{L, T}$) and the different lepton polarization asymmetries in the SM and in the THDM of type III.  The main results of the study can be summarized as
follows:
\begin{itemize}
\item We have observed that the longitudinal polarized branching ratio $\mathcal{B}_L$ deviate sizeably from the SM in different parametric space of the THDM type III.  In case of $\mu$.'s as final state leptons, the value of $\mathcal{B}_L$ differs by an order of magnitude from the SM values for different mass sets. The situation is same when we have calculated the transverse polarized branching ratio $(\mathcal{B}_T)$.

\item The longitudinal, normal and transverse polarization asymmetries of leptons are calculated for different parametric space of the THDM of type III in $B \to K^{\ast} \ell^+ \ell^-$ decays. The most promising shift comes in the range $s_{min}\leq s \leq m^2_{\psi}$ when we have $\mu$'s as final state leptons. However, in case of $\tau$'s the effects of the THDM are prominent in the range $14 GeV^2 \leq s \leq (m_B-m_{K^*})^2$. 

\end{itemize}
It is well known that experimentally it is hard to reconstruct the $\tau$'s, therefore, the most interesting decays are the ones which involve the $\mu$'s as final state leptons. In order to observe the longitudinal lepton polarization asymmetry in $B \to K^* \mu^+ \mu^-$ the number of events of $B\bar{B}$ required are around $10^{8} - 10^{10}$ which lies well in the range of LHC. Therefore, with the Run II of LHC, we hope that the measurement of these observables will help us to have some clues of the THDM.

\section*{Appendix}

\subsection*{Wilson Coefficients corresponding to NHB}
The exploit expressions of the Wilson coefficients corresponding to NHB contribution can be found in \cite{Huang} and at a scale $\mu = m_{W}$, these can be summarized as \cite{Falahati}:
\begin{eqnarray}
C_{Q_1}(m_{W})&=&\frac{m_bm_l}{m^2_{h^0}}\frac{1}{\left|\lambda_{tt}\right|^2}\frac{1}{sin^2\theta_{W}}\frac{x}{4}\big[(sin^2\alpha+hcos^2\alpha)f_1(x,y) \notag \\%
  &&+\left[\frac{m^2_{h^0}}{m^2_{W}}+(sin^2\alpha+hcos^2\alpha)(1-z)\right]f_2(x,y) \notag \\
&&+\frac{sin^2 2\alpha}{2m^2_{H^{\pm}}}\left[m^2_{h^0}-\frac{(m^2_{h^0}-m^2_{H^0})^2}{2m^2_{H^0}}\right]f_3(y)\big] \label{cq1} \\
C_{Q_2}(m_{W})&=&-\frac{m_bm_l}{m^2_{A^0}}\frac{1}{\left|\lambda_{tt}\right|^2}{f_1(x,y)+\big[1+\frac{m^2_{H^\pm}-m^2_{A^0}}{2m^2_{W}}\big]f_2(x,y)} \label{cq2} \\
C_{Q_3}(m_{W})&=&\frac{m_{b}e^2}{m_{\ell}g^2}\big[C_{Q_1}(m_{W})+C_{Q_2}(m_{W})\big] \label{cq3} \\
C_{Q_4}(m_{W})&=&\frac{m_{b}e^2}{m_{\ell}g^2}\big[C_{Q_1}(m_{W})-C_{Q_2}(m_{W})\big] \label{cq4} \\
C_{Q_i}(m_{W})&=&0 \enskip \enskip  i= 5,...,10 \label{cq5}
\end{eqnarray}
where
\begin{eqnarray}
x=\frac{m^2_t}{m^2_W}, y=\frac{m^2_t}{m^2_{H^\pm}}, z=\frac{x}{y}, h=\frac{m^2_{h^0}}{m^2_{H^0}}, \notag \\
f_1(x,y)=\frac{x\mathrm{ln}x}{x-1}-\frac{y\mathrm{ln}y}{y-1},\notag \\
 f_2(x,y)=\frac{x\mathrm{ln}y}{(z-x)(x-1)}-\frac{\mathrm{ln}z}{(z-1)(x-1)}, \notag \\
f_3(y)=\frac{1-y+y\mathrm{ln}y}{(y-1)^2}.
\end{eqnarray}
The evolution of the coefficients $C_{Q_1}$ and $C_{Q_2}$ is performed by the anomalous dimensions of $Q_{1}$ and $Q_2$, respectively:
\begin{equation}
C_{Q_i}(m_b)=\eta^{\gamma_Q/\beta_0}C_{Q_i}(m_W), \enskip \enskip i =1, 2
\end{equation}
where $\gamma_Q=-4$ is anomalous dimension of the operator $\bar{s}_Lb_R$.
 By using the knowledge of Wilson coefficients $C_7$, $\widetilde{C}_9$ and
$\widetilde{C}_{10}$ calculated at scale $m_W$, the Wilson coefficients $\widetilde{C}_{7}^{eff}$, $\widetilde{C}_{9}^{eff}$,
$\widetilde{C}_{10}$, $C_{Q_1}$ and $C_{Q_2}$ are calculated at the scale $m_{b}$. After adding the contribution from the charged Higgs
diagrams to the SM results, the Wilson coefficients $\widetilde{C}_{7}^{eff}$, $\widetilde{C}_{9}^{eff}$ and
$\widetilde{C}_{10}$ can take the form \cite{Falahati, Huang}:
\bea
\widetilde{C}_{7}(m_W)&=&C_{7}^{SM}(m_W)+\left|\lambda_{tt}\right|^2\left(\frac{y(7-5y-8y^2)}{72(y-1)^3}+\frac{y^2(3y-2)}{12(y-1)^4}\mathrm{ln}y\right) \nn \\
&&+\lambda_{tt}\lambda_{bb}\left(\frac{y(3-5y)}{12(y-1)^2}+\frac{y(3y-2)}{6(y-1)^3}\mathrm{ln}y\right), \label{C7mw} \\
\widetilde{C}_{9}(m_W)&=&\widetilde{C}_{9}^{SM}(m_W)+\left|\lambda_{tt}\right|^2\bigg[\frac{1-4sin^2\theta_W}{sin^2\theta_W}\frac{xy}{8}\left(\frac{1}{y-1}-\frac{1}{(y-1)^2}\mathrm{ln}y\right) \nn \\
&&-y\left(\frac{47y^2-79y+38}{108(y-1)^3}-\frac{3y^3-6y^2+4}{18(y-1)^4}\right)\bigg], \label{C9mw}\\
\widetilde{C}_{10}(m_W)&=&C_{10}^{SM}(m_W)+\left|\lambda_{tt}\right|^2\frac{1}{sin^2\theta_W}\frac{xy}{8}\left(-\frac{1}{y-1}+\frac{1}{(y-1)^2}\mathrm{ln}y\right)\label{C10mw}.
\eea

\subsection*{Longitudinal and Transverse Amplitudes for polarized branching ratio}
The amplitude for the longitudinally polarized branching ratio, i.e., $\mathcal{A}_L$ is:
\begin{eqnarray}
\mathcal{A}_L&=& 2(2m^2+q^2)(M_B^2-M^2_{K^{*}}-q^2)|f_2|^2+(4m^2(\lambda-12q^2M^2_{K^{*}})+2q^2(M_B^2-M^2_{K^{*}}-q^2)^2)|f_5|^2\notag \\
&&+4m^2\lambda(\lambda+3q^2(2M_B^2-q^2)+4\lambda q^2 M^2_{K^{*}})|f_6|^2+12m^2s^2\lambda|f_7|^2\notag \\
&&-(M_B^2-M^2_{K^{*}}-q^2)s(3\lambda-u^2(q^2))(f_2f_3^{*}+f_2^*f_3)\notag \\
&&-(12m^2\lambda s+(M_B^2-M^2_{K^{*}}-q^2)(3\lambda - u^2(q^2)))(f_5^*f_6+f_5f_6^*)\notag \\
&&+3s^2u^2(q^2)|f_8|^2-12m^2 s \lambda (f_7f_5^*+f_7^*f_5)\label{ALfunctions}
\end{eqnarray}
where as the transverse one $(\mathcal{A}_T)$ is
\begin{eqnarray}
\mathcal{A}_T(s)&=&(s+4m^2)\lambda |f_1|^2+2(2m^2+s)|f_2|^2+2(s-4m^2)(\lambda|f_4|^2+|f_5|^2) \notag \\
&&-\sqrt{\lambda}(s-4m^2)(f_4f^{*}_5-f_5f^*_4)\label{ATfunction}
\end{eqnarray}

\section*{Acknowledgments}
The author M. J. Aslam would like to thank Quaid-i-Azam University for the financial assistance through University Research Fund (URF).

\end{document}